\documentclass[12pt,a4paper,british]{iopart}

\usepackage[T1]{fontenc}
\usepackage[latin9]{inputenc}
\usepackage{babel}
\usepackage{graphicx}
\usepackage{epstopdf}
\usepackage[numbers,square,sort&compress]{natbib}
\usepackage[unicode=true,pdfusetitle,bookmarks=false,breaklinks=false,pdfborder={0 0 1},backref=false,colorlinks=false]{hyperref}
\usepackage{array}
\usepackage{verbatim}
\usepackage{amstext}
\usepackage{amssymb}
\usepackage{iopams}
\usepackage{setstack}
\usepackage{bibentry}

\usepackage{tocloft}

\makeatletter
\@ifundefined{definecolor}{\usepackage{color}}{}

\newcommand{\ket}[1]{| #1 \rangle}
\newcommand{\bra}[1]{\langle #1 |}
\newcommand{\braket}[2]{\langle #1|#2\rangle}
\newcommand{\ignore}[1]{}
\newcommand{\nobibentry}[1]{{\let\nocite\ignore\bibentry{#1}}}
\makeatother

\begin{document}

\title{Efficient tools for quantum metrology with uncorrelated noise}

\author{Jan Ko\l{}ody\'{n}ski and Rafa\l{} Demkowicz-Dobrza\'{n}ski}

\address{Faculty of Physics, University of Warsaw, 00-681 Warszawa, Poland}

\ead{jankolo@fuw.edu.pl \textrm{and} demko@fuw.edu.pl}

\nobibliography*
\begin{abstract}
Quantum metrology offers an enhanced performance in experiments such as gravitational wave-detection, magnetometry
or atomic clocks frequency calibration. The enhancement, however, requires
a delicate tuning of relevant quantum features such as entanglement or squeezing.
For any practical application the inevitable impact of decoherence needs to be taken into account in order
to correctly quantify the ultimate attainable gain in precision.
We compare the
applicability and the effectiveness of various methods of calculating the ultimate precision bounds resulting from
the presence of decoherence.
This allows us to put a number of seemingly unrelated concepts into a common framework
and arrive at an explicit hierarchy of quantum metrological methods
in terms of the tightness of the bounds they provide.
In particular, we show a way to extend the techniques
originally proposed in \nobibentry{Demkowicz2012},
so that they can be efficiently applied not only in the asymptotic
but also in the finite-number of particles regime.
As a result, we obtain a simple and direct method, yielding bounds that interpolate between
the quantum enhanced scaling characteristic for small number of particles and the asymptotic
regime, where quantum enhancement amounts
to a constant factor improvement.
Methods are applied to numerous models including noisy phase and frequency estimation,
as well as the estimation of the decoherence strength itself.
\end{abstract}

\pacs{03.67.-a, 03.65.Yz, 03.65.Ud, 03.65.Ta, 42.50.St}

%\submitto{\NJP}

\maketitle
%%%%%%%%%%%%%%%

{\pagestyle{plain}
\tableofcontents
\cleardoublepage}

%%%%%%%%%%%%%%%%%%%%%%%%%%%%%%%%%%%%%%%%%%%%%%%%%%%%%%%%%%%%%%%%%%%%%

\section{Introduction}

Quantum enhanced metrology has recently enjoyed a great success at experimental
level leading to new results in atomic spectroscopy \cite{Leibfried2004,Schmidt2005, Roos2006, Ospelkaus2011},
magnetometry \cite{Wasilewski2010,Wolfgramm2010,Koschorreck2010}
and optical interferometry \cite{Mitchell2004,Nagata2007,Resch2007,Higgins2009}
with prominent achievements in gravitational waves sensing \cite{LIGO2011}.
As predicted in the incipient theoretical results of \cite{Wineland1992,Bollinger1996,Buzek1999,Berry2000,Giovannetti2006},
a physical parameter unitarily encoded into a quantum system---a probe---consisting of $N$ entangled, non-interacting particles (atoms
or photons) can be extracted with a precision that is limited by the
quantum-mechanical uncertainty relations and not the more restrictive
central limit theorem of classical statistics. Hence, the uncertainty
in reconstructing the encoded parameter, such as e.g. optical phase delay or frequency difference, can in principle  be proportional
to $1/N$, the so-called \emph{Heisenberg Limit} (HL), rather than
$1/\sqrt{N}$, commonly referred to as the \emph{Standard Quantum
Limit} (SQL) or the shot (projection) noise. However, this dramatic
scaling improvement can be illusive, as both the experimental results
and theoretical toy-models have indicated that achieving the ideal
HL is a hard task owing to the strong destructive impact of imperfections,
which should be always accounted for in realistic scenarios.

\begin{figure}[!t]
\includegraphics[clip,width=0.95\textwidth]{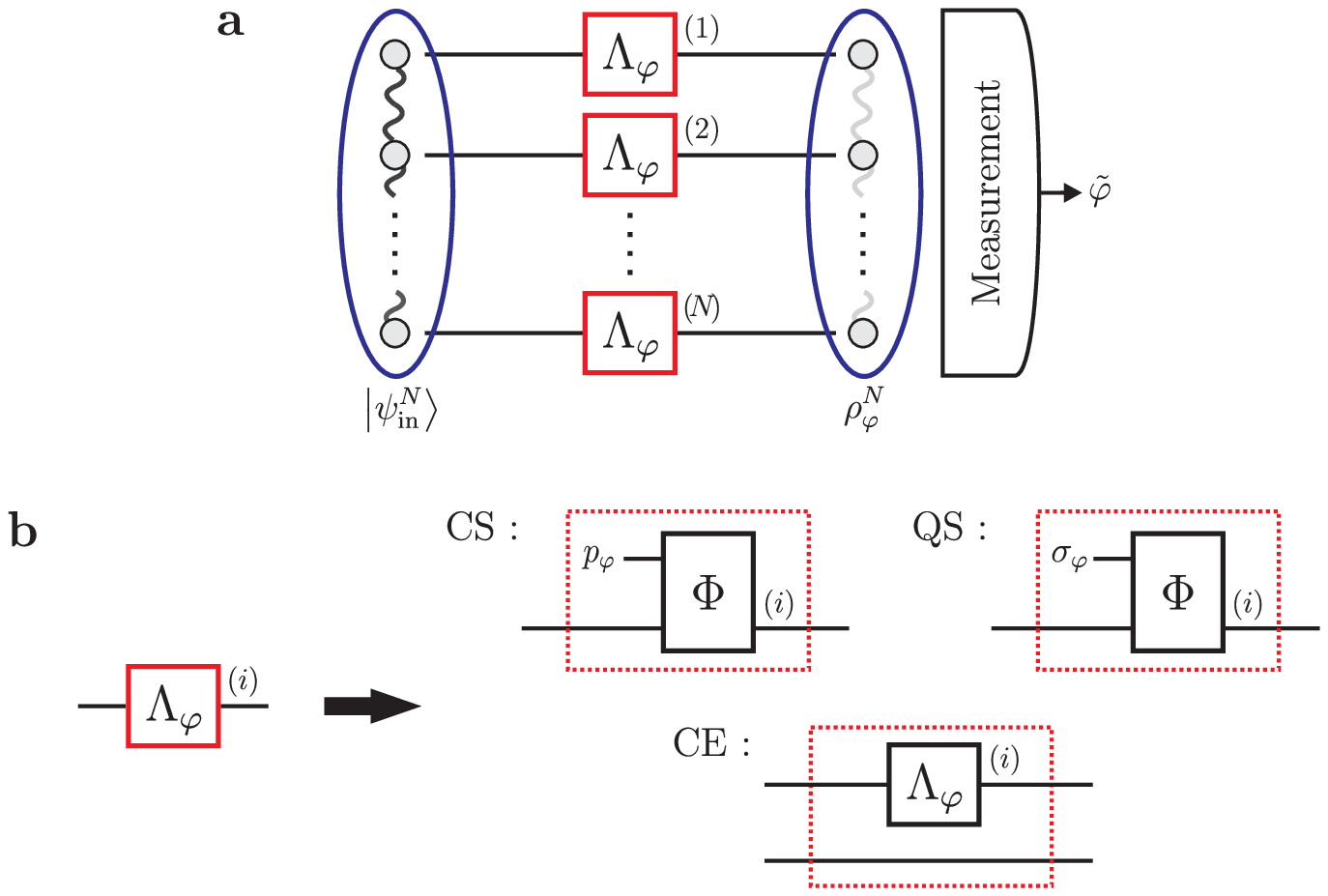}
\caption{\label{fig:qm_scheme}
\textbf{Quantum metrology and the single channel methods}
\protect \\
 (\textbf{a}) General scheme for quantum-enhanced metrology with uncorrelated
noise. The $N$ particles within the probe in a quantum state $\psi_{\textrm{in}}^{N}$
evolve and decohere independently while sensing an unknown parameter
$\varphi$ (e.g. phase). An estimate $\tilde{\varphi}$ of the parameter is inferred
from a measurement result on the final state of the probe $\rho_{\varphi}^{N}$.
\protect \\
(\textbf{b}) Precision bounds from single channel analysis.
The ultimate precision is bounded by the
$1/\sqrt{N}$ scaling (SQL), if for small variations $\delta\varphi$
around $\varphi$ the channel $\Lambda_{\varphi}$ can be expressed
as a parameter-independent map $\Phi$ that is also fed a classical,
diagonal state $p_{\varphi}$ -- CS, or a general, quantum state
$\sigma_{\varphi}$ -- QS, varying smoothly with $\varphi$. Still, for all such channels and more, the tightest bound on
precision is obtained by the CE method, in which the map $\Lambda_{\varphi}$
is replaced by its extension $\Lambda_{\varphi}\otimes\mathcal{I}$.}
\end{figure}

An important question that has been considerably addressed by many
researchers \cite{Banaszek2009,Giovannetti2011,Maccone2011,Dowling2008,Andre2004,Auzinsh2004,Huelga1997,
Shaji2007,Dorner2012,Huver2008,Meiser2009,Dorner2009,Demkowicz2009,Kolodynski2010,Knysh2011}
reads: how and to what extent can the noise effects be compensated in quantum metrological
setups?
%In the case of atomic spectroscopy, it has been shown that
%some dominant sources of error in implementations can be efficiently
%counterbalanced \cite{Andre2004,Auzinsh2004}, so that the SQL could
%have been significantly overcome in experiments such as \cite{Wasilewski2010,Wolfgramm2010,Koschorreck2010}.
%Yet, as indicated in \cite{Huelga1997,Shaji2007}, the effects of
%uncorrelated noise independently affecting atoms within the probe
%have a much more dramatic impact --- most likely restricting the ultimate
%precision scaling to become SQL-like for high enough $N$.
In the case of atomic spectroscopy, it has already been indicated in \cite{Huelga1997,Shaji2007} that the effects of
uncorrelated noise, independently affecting the atoms within a probe,
have a dramatic impact on quantum protocols---most likely restricting the ultimate
precision scaling to become SQL-like for high enough $N$, so that
the quantum enhancement is asymptotically limited to a multiplicative \emph{constant factor}.
%improvement and that the $1/\sqrt{N}$ scaling cannot be overcome for large $N$.
%More precisely, for a generic finite-strength uncorrelated environmental noise, acting
%independently on each of the probes, all quantum metrological schemes
%lead to an asymptotic precision formula of the form $c/\sqrt{N}$, were the $c$ factor
%depends on the strength and the type of the noise considered.
%Therefore, calculating the minimal achievable $c$ factors corresponds to
%determining the fundamental quantum precision enhancement bounds.
In optical interferometry, photonic loss is the main obstacle to practical implementations of quantum enhanced protocols
 \cite{Huver2008,Meiser2009,Dorner2009,Demkowicz2009} and the asymptotic SQL-like scaling is again inevitable, as proven
in \cite{Kolodynski2010,Knysh2011}.
%and experimentally studied in \cite{Kacprowicz2010}
Similarly to the atomic case, the asymptotic improvement constant
becomes an essential feature which determines the achievable precision
for high $N$. Following the above exemplary models, methods of quantifying
the asymptotic quantum enhancement have been proposed for arbitrary
kinds of probes with decoherence present \cite{Sarovar2006,Ji2008,Fujiwara2008,Matsumoto2010,Hayashi2011,Escher2011,Demkowicz2012}.
Recently, general procedures have been established that are capable of deriving practical
bounds on ultimate achievable precision in realistic quantum metrological setups
\cite{Escher2011, Demkowicz2012}. In particular, in the case of
uncorrelated noise it has been demonstrated that the sole analysis of the evolution of a \emph{single} particle often leads to
surprisingly informative bounds on the precision achievable with arbitrarily entangled multi-particle inputs 
\cite{Demkowicz2012}---see \Fref{fig:qm_scheme} for an outline of a relevant metrological scheme and a summary of the single particle (single channel)
methods investigated further on in this paper.
For completeness, it should be noted that there are some specific metrological models \emph{with} noise,
in which asymptotic scaling power enhancement \emph{is} nevertheless possible
\cite{Andre2004,Auzinsh2004, Chaves2012}.
Still, applicability of these models is limited, since in any practical implementation the noise types
considered will always be accompanied by
some generic decoherence processes for which the constant factor bound on the maximal quantum enhancement
will force the asymptotic precision scaling to be SQL-like.

In this paper, we provide new insight into relations between
seemingly unrelated methods used for derivations of various quantum metrological bounds and order them
with respect to their predictive power.
Firstly, focussing on the geometric intuitive method of channel \emph{classical
simulation }(CS) introduced in \cite{Matsumoto2010,Demkowicz2012},
we prove that its applicability is equivalent to the approach proposed
in \cite{Hayashi2011}. Moreover, we show that the criterion for
classical simulability of a channel can be generalized to a \emph{quantum
simulation} (QS) condition \cite{Matsumoto2010}, which coincides
with the channel programmability postulate of \cite{Ji2008}. Although
the idea of QS allows to prove the asymptotic SQL-like behaviour for
a wider range of decoherence models,
%including the lossy interferometer
we demonstrate that the \emph{channel extension} (CE) method of
\cite{Demkowicz2012} encompasses all CS, \cite{Hayashi2011}, QS
and \cite{Ji2008} approaches providing more stringent bounds on
precision. We also comment on the problem of the tightness of these bounds,
as they are guaranteed to be saturable only for channels, for which
the estimation task cannot be improved by allowing for an additional
ancilla (extension, idler mode) entangled with the input state. Classification
of such channels is an open problem of current research both in channel
estimation \cite{Acin2001,DAriano2001a,Chiribella2004,Hotta2005} and
channel discrimination \cite{Sacchi2005,DAriano2005a,Chiribella2008,Sedlak2009,Bisio2012,Pirandola2011,Pirandola2011a,Nair2011a,Nair2011}
theory. The graphical interpretation of the CS, QS and CE methods
is presented in \Fref{fig:qm_scheme}\textbf{b}.

Most importantly, we go beyond the results of \cite{Demkowicz2012}
and show that the CE method may be applied not only in the asymptotic regime but also
when dealing with finite probes of $N$ particles. Similarly to the
asymptotic case, it corresponds then to an optimisation procedure
over Kraus representations of a given channel that can be recast into
an efficiently solvable semi-definite programming task. We apply our
results to phase/frequency estimation with various noise models including: dephasing,
depolarization, loss and spontaneous emission, restricting
ourselves to the cases of noise commuting with the parameterized unitary
part of the evolution. This assumption  makes the analysis more transparent,
but is not indispensable, since our methods may be effectively employed for
any single particle evolution described by a general Lindblad equation
\cite{Breuer2002} reshaped into the corresponding Kraus representation
\cite{Andersson2007}. What is more, as our finite-$N$ CE method
applies to models for which its asymptotic version fails, it can be
used to upper-bound the asymptotic scaling for channels
surpassing the SQL. As an example, in \cite{Chaves2012}, our new method
has been already utilized to predict asymptotic super-classical scaling for
a channel with a non-commutative noise.
%Apart from the unitary evolution parameter estimation, other quantum channel characteristics may also
%be analyzed from the quantum metrological perspective
%\cite{Monras2007,Adesso2009,Crowley2012,Chiribella2006,Genoni2008,Aspachs2010,Latune2012}.
Additionally, in order to stress the generality of the methods, in the final section of this paper we show that
they can be applied not only to noisy unitary parameter estimation tasks but also to ones in which the decoherence strength itself is estimated.
%\cite{Monras2007,Adesso2009}.
Finally, we should also clarify that noise correlation and memory effects
\cite{Dorner2012, Matsuzaki2011,Chin2012,Szankowski2012} are beyond the scope of this paper; while the
non-Markovian effects, provided they affect each of the particles independently, might be analyzed using the
tools presented, correlations between the decoherence processes affecting the particles does not fit well into the framework advocated here.

This paper is organised as follows. In \Sref{sec:QFI} we introduce the mathematical
tools of estimation theory designed to quantify the achievable precision in metrological schemes
and discuss their applicability in the quantum setting. In \Sref{sec:QFICh} we study how these
concepts may be utilized when the estimated parameter is encoded during the evolution of a given
quantum system. \Sref{sec:QFIChN} is devoted to quantum systems that consist of $N$ particles
undergoing independent evolution and contains the main results of the paper---methods allowing to quantify
the ultimate precision both in the finite-$N$ and asymptotic $N$ regimes as well as their direct application
to phase estimation schemes.  In \Sref{sec:FreqEstim} we show how these methods should be accommodated
in order to encompass the frequency estimation tasks of atomic spectroscopy, whereas in \Sref{sec:LossEstim}
when considering metrological scenarios in which the strength of noise or loss is the parameter to be estimated.
\Sref{sec:FurDisc} contains additional discussion on consequences of the
results obtained as well as an outlook on future research.
\Sref{sec:SumOut} summarizes the paper.

%%%%%%%%%%%%%%%%%%%%%%%%%%%%%%%%%%%%%%%%%%%%%%%%%%%%%%%%%%%%%%%%%%%%%%%%%%%%%%%%%%%%%%%%%%%5

\section{Quantum Fisher Information \label{sec:QFI}}

\subsection{Classical Cram\'{e}r-Rao bound}

Let us assume that after measuring a physical system an outcome is
obtained that can be represented by a random variable $X$ distributed
with some probability distribution $p_{\varphi}(X)$. If the system
is classically described, all its properties can be simultaneously
determined, so that $X$ can in principle be multidimensional and
contain as much information about the system as allowed by the available
resources. The estimation task corresponds then to determining with
highest precision the quantity $\varphi$ based on the observed value of $X$.
As stated by the
\emph{Cram\'{e}r-Rao bound} \cite{Kay1993} any unbiased strategy
to determine the unknown parameter after repeating the procedure $k$
times, must provide an estimate $\tilde{\varphi}$ with uncertainty
that is lower bounded by
\begin{equation}
\Delta\tilde{\varphi}\ge\frac{1}{\sqrt{k\, F_{\textrm{cl}}\!\left[p_{\varphi}\right]}}\textrm{ ,\qquad\quad where\qquad}F_{\textrm{cl}}\!\left[p_{\varphi}\right]=\int\!\! dx\;
\frac{\,{\dot{p}_{\varphi}(x)}^{2}}{p_{\varphi}(x)}\label{eq:clCRB}
\end{equation}
is the (\emph{classical})\emph{ Fisher Information} (FI) \footnote{Throughout the paper, we depict derivatives w.r.t. the estimated parameter with an `overdot', so that e.g. $\dot{p}_\varphi(x) \equiv \partial_{\varphi}\,p_{\varphi}(x)$, $\dot{\rho}_\varphi \equiv \partial_{\varphi}\, \rho_\varphi$ and $\dot{K}(\varphi) \equiv \partial_{\varphi}\, K(\varphi)$.}.

The $1/\sqrt{k}$ dependence in \eref{eq:clCRB} is a consequence
of the central limit theorem and the fact that the $k$ procedures are
independent. This manifests itself by the additivity property of the
FI, i.e. $F_{\textrm{cl}}\!\left[p_{\varphi}^{k}\right]\!=\! k\, F_{\textrm{cl}}\!\left[p_{\varphi}\right]$
for $X^{k}$.
%The FI can be interpreted as a distance measure between
%neighbouring probability distributions $p_{\varphi}$ and $p_{\varphi+\delta\varphi}$.
%Defining a statistical distance as an angle between two distributions
%$p$ and $q$, i.e. $D\!\left(p,q\right)\!=\!\arccos(F(p,q))$ with
%fidelity $F(p,q)\!=\!\int\!\! dx\sqrt{p(x)\, q(x)}$ \cite{Nielsen2000},
%we can Taylor expand it for small deviations $\delta\varphi\!>\!0$
%to obtain the so called FI distance \cite{Bengtsson2006}
%\begin{equation}
%D\!\left(p_{\varphi},p_{\varphi+\delta\varphi}\right)=\frac{1}{2}\sqrt{F_{\textrm{cl}}\!\left[p_{\varphi}\right]}\,\delta\varphi+O\!\left(\delta\varphi^{2}\right).\label{eq:FIviaStDist}
%\end{equation}
\Eref{eq:clCRB} shows that the FI is a \emph{local }quantity containing
information about infinitesimal variations of $\varphi$. That is
why, FI is designed to be used in the so called \emph{local estimation} approach in which small
parameter fluctuations are to be sensed. This small deviations regime may always be
reached after many procedure repetitions ($k\!\rightarrow\!\infty$)
and in this limit the Cram\'{e}r-Rao bound is known to be saturable via e.g. max-likelihood estimation schemes \cite{Kay1993}.

\subsection{Quantum Cram\'{e}r-Rao bound}

In a quantum estimation scenario, the parameter $\varphi$ is encoded in a quantum state $\rho_\varphi$.
A general measurement,  mathematically represented by the elements of the \emph{positive operator
valued measure }(POVMs) $M_{x}$ that satisfy $M_x \geq 0$, $\int\! dx\, M_{x}\!=\!\mathbb{I}$
\footnote{We denote by $\mathbb{I}$---the identity operator
and by $\mathcal{I}$---the identity superoperator.} \cite{Nielsen2000,Bengtsson2006},
 is performed yielding outcome statistics  $p_{\varphi}(X)\!=\!\textrm{Tr}\left\{ \rho_{\varphi}M_{X}\right\}$.
Establishing the optimal
estimation strategy corresponds then not only to the correct interpretation of
the measurement results, but also to a non-trivial optimization over
the class of all POVMs to find the measurement scheme maximizing the
precision. In this case the \emph{quantum Cram\'{e}r-Rao
bound} can be derived \cite{Helstrom1976, Holevo1982, Braunstein1994}, which is independent of the choice of the POVMs
and solely determined by the dependence of the output state on the
estimated parameter:
\begin{equation}
\Delta\tilde{\varphi}\ge\frac{1}{\sqrt{k\, F_{\textrm{Q}}\!\left[\rho_{\varphi}\right]}}\textrm{ \qquad\quad with\qquad\quad}F_{\textrm{Q}}\!\left[\rho_{\varphi}\right]=\textrm{Tr}\!\left\{ \rho_{\varphi}\left.L^{\textrm{S}}[\rho_{\varphi}]\right.^{2}\right\} \label{eq:qCRB}
\end{equation}
being now the \emph{quantum Fisher information} (QFI). The Hermitian
operator $L^{\textrm{S}}[\rho_{\varphi}]$ is the so called \emph{symmetric
logarithmic derivative} (SLD), which can be unambiguously defined
for any state $\rho_{\varphi}$ via the relation $\dot{\rho}_{\varphi}\!=\!\frac{1}{2}\left(\rho_{\varphi}L^{\textrm{S}}[\rho_{\varphi}]\!+\! L^{\textrm{S}}[\rho_{\varphi}]\rho_{\varphi}\right)$.
Then, in the eigenbasis of $\rho_{\varphi}\!=\!\sum_{i}\lambda_{i}(\varphi)\left|e_{i}(\varphi)\right\rangle \!\left\langle e_{i}(\varphi)\right|$
with $\left\{ \left|e_{i}(\varphi)\right\rangle \right\} _{i}$ forming
a complete basis ($\forall_{i}\!:\,0\!\le\!\lambda_{i}\!\le\!1$)
\begin{equation}
L^{\textrm{S}}[\rho_{\varphi}]=\sum_{i,j}\frac{2\left<e_{i}\!\left(\varphi\right)
\right|\dot{\rho}_{\varphi}\left|e_{j}\!\left(\varphi\right)\right>}{\lambda_{i}
\!\left(\varphi\right)+\lambda_{j}\!\left(\varphi\right)}\left|e_{i}\!\left(\varphi\right)\right>\!\left<e_{j}\!\left(\varphi\right)\right|,
\label{eq:SLD}
\end{equation}
where the sum is taken over the terms with non-vanishing denominator.
QFI is an additive quantity on product states and in particular $F_{\textrm{Q}}\!\left[\rho_{\varphi}^{\otimes k}\right]\!=\! k\,
F_{\textrm{Q}}\!\left[\rho_{\varphi}\right]$. Thus, the $\sqrt{k}$ term in the denominator of \eref{eq:qCRB} may be equivalently interpreted as
the number of independent repetitions of an experiment with a state $\rho_\varphi$ or a single shot experiment
with a multi-party state $\rho_\varphi^{\otimes k}$.
Crucially, as proven in \cite{Braunstein1994,Nagaoka2005}, there
always exist a measurement strategy, e.g. projection measurement in the eigenbasis of the SLD, for which bounds
\eref{eq:clCRB} and \eref{eq:qCRB} coincide. Hence,
as in the classical case, the saturability of \eref{eq:qCRB} is
guaranteed, but again only in the $k\!\rightarrow\!\infty$ limit.

\subsection{Purification-based definition of QFI}

For pure states, $\rho_{\varphi}\!=\!\left|\psi_{\varphi}\right\rangle \!\left\langle \psi_{\varphi}\right|$,
the QFI in \eref{eq:qCRB} simplifies to $F_{\textrm{Q}}\!\left[\left|\psi_{\varphi}\right\rangle \right]\!=\!4\left(\left\langle \dot{\psi}_{\varphi}|\dot{\psi}_{\varphi}\right\rangle \!-\!\left|\left\langle \dot{\psi}_{\varphi}|\psi_{\varphi}\right\rangle \right|^{2}\right)$
\footnote{We shorten the notation of functions and superoperators
of pure states, so that $F\!\left[\left|\psi\right\rangle \right]\equiv F\!\left[\left|\psi\right\rangle \!\left\langle \psi\right|\right]$
and $\Lambda\!\left[\left|\psi\right\rangle \right]\equiv\Lambda\!\left[\left|\psi\right\rangle \!\left\langle \psi\right|\right]$.}.
Yet, as indicated by \eref{eq:SLD}, for general mixed states the
computation of QFI involves diagonalisation of $\rho_{\varphi}$, which
may be infeasible when dealing with large systems. That is why it
is often necessary to look for upper bounds on QFI that would be efficiently
calculable even at the expense of saturability. For this purpose, definitions
of QFI were proposed that do not involve computing the SLD, but are
specified at the level of state purifications: $\rho_{\varphi}\!=\!\textrm{Tr}_{\textrm{E}}\left\{ \left|\Psi(\varphi)\right\rangle \!\left\langle \Psi(\varphi)\right|\right\} $.
In \cite{Escher2011} the QFI of any $\rho_{\varphi}$ has been proven
to be equal to the smallest QFI of its purifications $\left|\Psi(\varphi)\right\rangle $
\footnote{See also an alternative formulation based on the convex roof formula, which is valid for unitary parameter estimation
\cite{Toth2013, Yu2013}.}:
\begin{equation}
\!\!\!\!\!\!\!\!\! F_{\textrm{Q}}\!\left[\rho_{\varphi}\right]=\min_{\Psi(\varphi)}F_{\textrm{Q}}\!\left[\left|\Psi(\varphi)\right\rangle \right]=4\min_{\Psi(\varphi)}\left\{ \left\langle \dot{\Psi}(\varphi)|\dot{\Psi}(\varphi)\right\rangle -\left|\left\langle \dot{\Psi}(\varphi)|\Psi(\varphi)\right\rangle \right|^{2}\right\} \!.\label{eq:FqPurifEscher}
\end{equation}
Independently, in \cite{Fujiwara2008} another purification-based QFI definition has been constructed:
\begin{equation}
\!\!\!\!\!\!\!\!\! F_{\textrm{Q}}\!\left[\rho_{\varphi}\right]=4\min_{\Psi(\varphi)}\left\langle \dot{\Psi}(\varphi)|\dot{\Psi}(\varphi)\right\rangle \!.\label{eq:FqPurifFuji}
\end{equation}
Despite apparent difference, Eqs.~\eref{eq:FqPurifEscher} and \eref{eq:FqPurifFuji}
are equivalent and one can prove that any purification
minimizing one of them is likewise optimal for the other and satisfies
the condition $\left|\dot{\Psi}(\varphi)\right\rangle \!=\!\frac{1}{2}L^{\textrm{S}}\!\left[\rho_{\varphi}\right]\!\otimes\!\mathbb{I}_{\textrm{E}}\left|\Psi(\varphi)\right\rangle $
causing the second term of \eref{eq:FqPurifEscher} to vanish. Although
for any suboptimal $\Psi(\varphi)$ \eref{eq:FqPurifEscher} must
provide a strictly tighter bound on QFI than \eref{eq:FqPurifFuji},
the latter definition, owing to its elegant form, allows for more
agility in the minimization procedure, so that it has been efficiently
utilized in \cite{Fujiwara2008,Demkowicz2012} and is also the base for this paper.

\subsection{Right logarithmic derivative (RLD)-based upper bound on QFI}
\label{sub:FqRLD}

On the other hand, a natural way to construct a bound on QFI and avoid
the SLD computation is to relax the Hermiticity condition of the logarithmic
derivative. If a non-Hermitian $L[\rho_{\varphi}]$ satisfying $\partial_{\varphi}\rho_{\varphi}\!=\!\frac{1}{2}\left(\rho_{\varphi}L[\rho_{\varphi}]\!+\! L[\rho_{\varphi}]^{\dagger}\rho_{\varphi}\right)$
can be found, as proven in \cite{Holevo1982, Nagaoka2005}, an upper limit on
QFI in \eref{eq:qCRB} is obtained: $F_{\textrm{Q}}\!\left[\rho_{\varphi}\right]\!\le\!\textrm{Tr}\!\left\{ \rho_{\varphi}L[\rho_{\varphi}]L[\rho_{\varphi}]^{\dagger}\right\} $.
In particular, if and only if $\partial_{\varphi}\rho_{\varphi}$
is contained within the support of $\rho_{\varphi}$, one can construct
the \emph{Right Logarithmic Derivative }(RLD) by setting $L[\rho_{\varphi}]\!=\! L^{\textrm{R}}[\rho_{\varphi}]\!=\!\rho_{\varphi}^{-1}\,\partial_{\varphi}\rho_{\varphi}$
and formulate an upper bound on QFI of a simpler form:
\begin{equation}
F_{\textrm{Q}}\!\left[\rho_{\varphi}\right]\le F_{\textrm{Q}}^{\textrm{RLD}}\!\left[\rho_{\varphi}\right]=\textrm{Tr}\!\left\{ \rho_{\varphi}^{-1}\,\left(\partial_{\varphi}\rho_{\varphi}\right)^{2}\right\} \!.\label{eq:FqRLD}
\end{equation}
Although
% in the case of RLD,
\eref{eq:FqRLD} is tight only when $L^{\textrm{R}}[\rho_{\varphi}]\!=\! L^{\textrm{S}}[\rho_{\varphi}]$,
it still allows one to quantify precision well for channel estimation
tasks \cite{Hayashi2011}, as described in the following section.
Lastly, one should note that we are not considering here multi-parameter
estimation schemes, for which the RLD may sometimes provide tighter
bounds than the SLD \cite{Fujiwara1994,Genoni2012}.

%%%%%%%%%%%%%%%%%%%%%%%%%%%%%%%%%%%%%%%%%%%%%%%%%%%%%%%%%%%%%%%%%%%%%%%%%%%%%%%%%%%%%%%%%%%5

\section{Estimation of a quantum channel \label{sec:QFICh}}

\subsection{Channel QFI}

As in metrological setups the estimated parameter is encoded in the
evolution of a system, we identify $\rho_{\varphi}\!=\!\Lambda_{\varphi}[\rho_{\textrm{in}}]$
as the final state of a system that started from an input $\rho_{\textrm{in}}$.
The preparation of $\rho_{\textrm{in}}$ is controlled in order to
achieve the most precise estimate of $\varphi$ that parametrizes
some general \emph{channel}---a \emph{Completely Positive Trace
Preserving} (CPTP) map $\Lambda_{\varphi}$ \cite{Nielsen2000,Bengtsson2006}.
Although the form of $\Lambda_{\varphi}$ in general strongly depends
on the model considered, the Quantum Cram\'{e}r-Rao bound\emph{ }always
applies, so that precision is upper bounded according to \eref{eq:qCRB}
with QFI $F_{\textrm{Q}}\!\left[\Lambda_{\varphi}[\rho_{\textrm{in}}]\right]$.
Furthermore, as the QFI is a convex quantity \cite{Fujiwara2001},
one should restrict oneself to pure input states when seeking the
optimal one.
%yielding the highest precision.
Hence, as shown in \Fref{fig:ch_purif_ext}\textbf{a}, we define
the \emph{channel QFI }as the maximal QFI after performing the input
optimisation, so that it has a concrete operational and application-like
interpretation
\begin{equation}
\mathcal{F}\!\left[\Lambda_{\varphi}\right]=\max_{\psi_{\textrm{in}}}\, F_{\textrm{Q}}\!\left[\,\Lambda_{\varphi}\!\left[\left|\psi_{\textrm{in}}\right\rangle \right]\right]\!.\label{eq:FqCh}
\end{equation}

For instance, while estimating the duration of the evolution ($\varphi\!\equiv\! t$)
in an ideal, decoherence-free setting, the CPTP map is unitary leading
to a pure channel output. Thus, $\Lambda_{\varphi}\!\left[\left|\psi_{\textrm{in}}\right\rangle \right]\!=\!\mathcal{U}_{t}\!\left[\left|\psi_{\textrm{in}}\right\rangle \right]\!=\!\textrm{e}^{-\textrm{i}Ht}\left|\psi_{\textrm{in}}\right\rangle \!\left\langle \psi_{\textrm{in}}\right|\textrm{e}^{\textrm{i}Ht}$
with $H$ being the Hamiltonian of the evolution. Hence, the definition
\eref{eq:FqCh} corresponds to \begin{equation}
\!\!\!\!\!\!\!\!\!\!\!\!\!\!\!\!\!\!\!\!\!\!\!\!\!\!\!\!\!\!\!\!\!\!\mathcal{F}\!\left[\,\mathcal{U}_{t}\right]=\max_{\psi_{\textrm{in}}}\,
F_{\textrm{Q}}\!\left[\textrm{e}^{-\textrm{i}Ht}\!\left|\psi_{\textrm{in}}\right\rangle \right]=4\max_{\psi_{\textrm{in}}}\!\left\{ \left\langle
 \psi_{\textrm{in}}\right|\!H^{2}\!\left|\psi_{\textrm{in}}\right\rangle -\left\langle
 \psi_{\textrm{in}}\right|\!H\!\left|\psi_{\textrm{in}}\right\rangle ^{2}\right\} =4 \max_{\psi_{\textrm{in}}}\,\Delta^{2}H,\label{eq:FqChUnit}
\end{equation}
and the optimal states are the ones that maximize the Hamiltonian variance.
Note also that in this case the quantum Cram\'{e}r-Rao bound takes the form of the
time-energy uncertainty relation, $\Delta H\!\cdot\!\Delta \tilde{t}\!\ge\!1/2$
\cite{Braunstein1996,Aharonov2002}, with $\Delta \tilde{t}$ being the uncertainty in the estimated duration.

\subsection{Purification-based definition of channel QFI}

\begin{figure}[!t]
\includegraphics[width=1\textwidth]{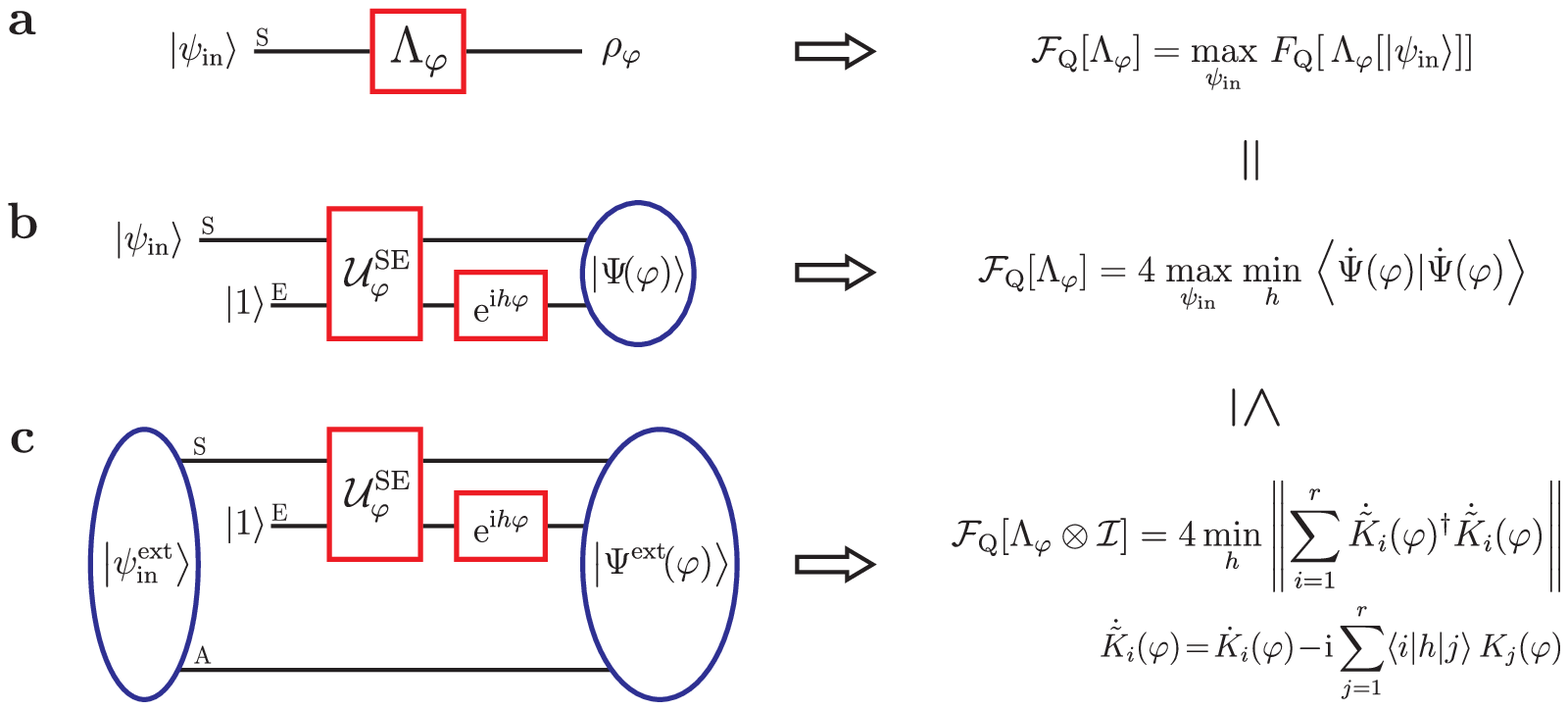}
\caption{
\label{fig:ch_purif_ext}
\textbf{Channel QFI based on the output state purification}
\protect \\
(\textbf{a}) Channel QFI as the QFI of the output
state maximized over all pure input states.
\protect \\
(\textbf{b}) Channel QFI obtained from the output state purification
generated by a local, fictitious, parameter-dependent environment
rotation.
\protect \\
(\textbf{c}) Extended channel QFI independent of the maximization
over the input states. The environment rotation corresponds to a
choice of Kraus representation of the channel.}
\end{figure}

In order to employ the definition \eref{eq:FqPurifFuji}, we utilize
the Stinespring theorem \cite{Bengtsson2006} and express the channel
$\Lambda_{\varphi}$ as a unitary map $U_{\varphi}^{\textrm{SE}}$
on the system combined with an environment disregarded after the evolution.
In this way, the output state and its purification respectively read
$\Lambda_{\varphi}\!\left[\left|\psi_{\textrm{in}}\right\rangle \right]\!=\!\textrm{Tr}_{\textrm{E}}\!\left\{ \left|\Psi(\varphi)\right\rangle \!\left\langle \Psi(\varphi)\right|\right\} $
and $\left|\Psi(\varphi)\right\rangle \!=\! U_{\varphi}^{\textrm{SE}}\left|\psi_{\textrm{in}}\right\rangle \!\otimes\!\left|1\right\rangle $,
where $\left|1\right\rangle $ is an arbitrary fixed state chosen
to be the first vector in the basis $\left\{ \left|i\right\rangle \right\} _{i=1}^{r}$
of the environment Hilbert space $\mathcal{H}_{\textrm{E}}^{r}$.
By specifying the dimension $r$ of $\mathcal{H}_{\textrm{E}}^{r}$
to be equal to the rank of $\Lambda_{\varphi}$, we can generate all
non-trivial purifications, $\tilde{\Psi}(\varphi)$, by applying a
fictitious, possibly $\varphi$-dependent unitary rotation, $u_{\varphi}^{\textrm{E}}$,
to the environment, so that $\left|\tilde{\Psi}(\varphi)\right\rangle \!=\!\tilde{U}_{\varphi}^{\textrm{SE}}\left|\psi_{\textrm{in}}\right\rangle \!\otimes\!\left|1\right\rangle $
with $\tilde{U}_{\varphi}^{\textrm{SE}}\!=\! u_{\varphi}^{\textrm{E}}\, U_{\varphi}^{\textrm{SE}}$.
Furthermore, writing the channel action in its Kraus representation
form $\Lambda_{\varphi}\!\left[\left|\psi_{\textrm{in}}\right\rangle \right]\!=\!\sum_{i=1}^{r}\tilde{K}_{i}(\varphi)\left|\psi_{\textrm{in}}\right\rangle \!\left\langle \psi_{\textrm{in}}\right|\tilde{K}_{i}(\varphi)^{\dagger}$,
we can identify the Kraus operators corresponding to $\tilde{\Psi}(\varphi)$
as
\begin{equation}
\tilde{K}_{i}(\varphi)=\left\langle i\right|\tilde{U}_{\varphi}^{\textrm{SE}}\left|1\right\rangle =
\sum_{j=1}^{r}u^{\textrm{E}}_{\varphi,ij}\, K_{j}(\varphi),\label{eq:KrausReps}
\end{equation}
where $u^{\textrm{E}}_{\varphi,ij}\!=\!\left\langle i\right|u_{\varphi}^{\textrm{E}}\left|j\right\rangle $
and $K_{j}(\varphi)\!=\!\left\langle j\right|U_{\varphi}^{\textrm{SE}}\left|1\right\rangle $
are the Kraus operators of the original purification $\Psi(\varphi)$.
%\footnote{We use the bold font to indicate that $\mathbf{u}(\varphi)$ and $\mathbf{\mathbf{h}}$
%should be treated as complex matrices and not operators, despite having
%same entries as $u_{\varphi}^{\textrm{E}}$ and $h^{\textrm{E}}$
%respectively.}.
Hence, by picking an environment unitary rotation $u_{\varphi}^{\textrm{E}}$,
we are, equivalently to the purification choice, specifying a Kraus
representation of $\Lambda_{\varphi}$. Moreover, as the QFI is a
local quantity, we can restrict ourselves
to infinitesimal rotations, $u_{\varphi}^{\textrm{E}}\!=\!\textrm{e}^{-\textrm{i}h (\varphi-\varphi_{0})}$,
in the vicinity of the real value $\varphi_{0}$ that are generated
by some Hermitian $h$. Taking without loss of generality
$\varphi_{0}\!=\!0$, we obtain the channel version of \eref{eq:FqPurifFuji}
shown in \Fref{fig:ch_purif_ext}\textbf{b} as
\begin{equation}
\label{eq:maxmin}
\!\!\!\!\!\!\!\!\!\!\!\!\!\!\!\!\!\!\!\!\!\!\mathcal{F}\!\left[\Lambda_{\varphi}\right]  =
4\max_{\psi_{\textrm{in}}}\,\min_{h}\;\braket{\dot{\tilde{\Psi}}(\varphi)}{\dot{\tilde{\Psi}}(\varphi)} =
%\textrm{Tr}_{\textrm{S}}\!\left\{ \left|\psi_{\textrm{in}}\right\rangle \!\left\langle \psi_{\textrm{in}}\right|\;\left\langle %1\right|\partial_{\varphi}\tilde{U}_{\varphi}^{\textrm{SE}\dagger}\,\partial_{\varphi}\tilde{U}_{\varphi}^{\textrm{SE}}\left|1\right\rangle \right\}
 4\max_{\psi_{\textrm{in}}}\,\min_{h}\; \bra{\psi_{\textrm{in}}}\! \sum_{i=1}^{r}\dot{\tilde{K}}_{i}(\varphi)^{\dagger}\dot{\tilde{K}}_{i}(\varphi) \ket{\psi_{\textrm{in}}} ,\label{eq:FqChPurif}
\end{equation}
where
$\ket{\dot{\tilde{\Psi}}(\varphi)}\!=\!(\dot{U}^{\textrm{SE}}_\varphi - \textrm{i} h U^{\textrm{SE}}_\varphi) \ket{\psi_{\textrm{in}}} \otimes \ket{1}$
and similarly $\dot{\tilde{K}}_{i}(\varphi)\!=\!\dot{K}_{i}(\varphi)-\textrm{i}\sum_{j=1}^{r}h_{ij}K_{j}(\varphi)$
with $h_{ij}\!=\!\left\langle i\right|h \left|j\right\rangle $
being the elements of the generator of Kraus representation rotations
\eref{eq:KrausReps}.
Crucially, \Fref{fig:ch_purif_ext}\textbf{b} and \Eref{eq:FqChPurif}
indicate that the optimal purification/Kraus representation corresponds
to the choice of an artificial environment that rotates locally with
$\varphi$ hindering as much as possible information about the estimated
parameter.

In order to make the reader familiar with the above formalism, we apply the definition \eref{eq:FqChPurif} to the
previously mentioned case of the evolution duration estimation $\mathcal{U}_{t}[\left|\psi_{\textrm{in}}\right\rangle ]$.
As the evolution is unitary and $r\!=\!1$, the environment/Kraus
rotations correspond just to a phase variation $u_{t}^{\textrm{E}}\!=\!\textrm{e}^{-\textrm{i}h\,t}$
with $h$ being a real scalar. Hence, \eref{eq:FqChPurif} simplifies
to \eref{eq:FqChUnit} as expected:
\begin{equation}
\!\!\!\!\!\!\!\!\!\!\!\!\!\!\!\!\!\!\!\mathcal{F}\!\left[\,\mathcal{U}_{t}\right]=4\max_{\psi_{\textrm{in}}}\min_{h}\!\left\{
\left\langle \psi_{\textrm{in}}\right|\!H^{2}\!\left|\psi_{\textrm{in}}\right\rangle -2h\left\langle \psi_{\textrm{in}}
\right|\!H\!\left|\psi_{\textrm{in}}\right\rangle +h^{2}\right\} =4\max_{\psi_{\textrm{in}}}\,\Delta^{2}H
\end{equation}
with minimum occurring at $h\!=\!\left\langle \psi_{\textrm{in}}\right|\!H\!\left|\psi_{\textrm{in}}\right\rangle $.
Consistently, we would also arrive at \eref{eq:FqChUnit}, if we had
used the other QFI purification-based definition \eref{eq:FqPurifEscher}
as shown in \cite{Escher2011}.

\subsection{Extended channel QFI}
\label{sub:QFIChExt}

The channel QFI \eref{eq:FqCh} is
affected by any $\varphi$-variations in the form of $\Lambda_{\varphi}$,
quantifying the distinguishability between maps $\Lambda_{\varphi}$
and $\Lambda_{\varphi+\delta\varphi}$. However, any such disturbance
may be noticeable only for input states which lead to a measurable
change of the channel output that is at best in some ``orthogonal
direction''. As a consequence, the quantity $\min_{h} \left\{ \dots\right\} $
in \eref{eq:FqChPurif} depends strongly on the pure input $\psi_{\textrm{in}}$,
as the minimum occurs for Kraus operators $\left\{ K_{i}^{\textrm{opt}}\!(\varphi)\right\} _{i=1}^{r}$
which fulfill the condition $\dot{K}_{i}^{\textrm{opt}}\!(\varphi)\left|\psi_{\textrm{in}}\right\rangle \!=\!\frac{1}{2}L^{\textrm{S}}\!\left[\Lambda_{\varphi}\left[\left|\psi_{\textrm{in}}\right\rangle \right]\right]\, K_{i}^{\textrm{opt}}\!(\varphi)\left|\psi_{\textrm{in}}\right\rangle $
required for the purification of \eref{eq:FqPurifEscher} and \eref{eq:FqPurifFuji}
to be optimal. Maximization of this quantity over $\ket{\psi_{\textrm{in}}}$ is difficult in general, due to the impossibility
of exchanging the order of $\max$ and $\min$ in \eref{eq:maxmin} \cite{Fujiwara2008}.

Yet, one may construct a natural upper bound on the channel QFI \eref{eq:FqCh}
by \emph{extending }the input space, $\mathcal{H}_{\textrm{S}}$,
by an equally-large auxiliary space, $\mathcal{H}_{\textrm{A}}$,
which is unaffected by the map and measured along with the channel
output (see \Fref{fig:ch_purif_ext}\textbf{c}). This way, by employing
extended input states entangled between these two spaces, $\left|\psi_{\textrm{in}}^{\textrm{ext}}\right\rangle \!\in\!\mathcal{H}_{\textrm{S}}\!\times\!\mathcal{H}_{\textrm{A}}$,
one may acquire full available information about $\varphi$ imprinted
by the map $\Lambda_{\varphi}$ on the extended output state. The
analogue of \eref{eq:FqChPurif} defines then the \emph{extended channel
QFI}\cite{Fujiwara2008}:
\begin{equation}
\!\!\!\!\!\!\!\!\!\!\!\!\!\!\!\!\!\!\!\!\!\!\!\!\!\!\!\!\!
\mathcal{F}\!\left[\Lambda_{\varphi}\otimes\mathcal{I}\right]=4\max_{\rho_{\textrm{in}}^{\textrm{S}}}\min_{h}\,
\textrm{Tr}_{\textrm{S}}\!\left\{ \rho_{\textrm{in}}^{\textrm{S}}\sum_{i=1}^{r}\!\dot{\tilde{K}}_{i}(\varphi)^{\dagger}
\dot{\tilde{K}}_{i}(\varphi)\right\} \!=4\min_{h}\left\Vert \sum_{i=1}^{r}\!\dot{\tilde{K}}_{i}(\varphi)^{\dagger}
\dot{\tilde{K}}_{i}(\varphi)\right\Vert \!,\label{eq:FqChExt}
\end{equation}
where $\left\Vert \dots\right\Vert $ represents the operator norm.
The first expression is obtained by tracing over the auxiliary space
$\mathcal{H}_{\textrm{A}}$, what leads to the maximisation over all
mixed states
$\rho_{\textrm{in}}^{\textrm{S}}\!=\!\textrm{Tr}_{\textrm{A}}\!\left\{ \left|\psi_{\textrm{in}}^{\textrm{ext}}\right\rangle \!\left\langle
\psi_{\textrm{in}}^{\textrm{ext}}\right|\right\} $.
However, this is exactly \eref{eq:FqChPurif} with the pure input
state replaced by a general mixed one, in which case the order of $\max$ and $\min$ can be swapped \cite{Fujiwara2008}
\footnote{We should stress that \eref{eq:FqChExt} \emph{does not} correspond to
the situation of using \emph{mixed} states as inputs for \emph{unextended} channel, as mixed states never outperform pure state inputs
due to convexity of the QFI.}.
Consistently,
if the optimal input state of \eref{eq:FqChExt} is pure, \eref{eq:maxmin}
and \eref{eq:FqChExt} become equivalent manifesting the uselessness
of entanglement between $\mathcal{H}_{\textrm{S}}$ and $\mathcal{H}_{\textrm{A}}$
and the irrelevance of the auxiliary space.

Importantly, the extended channel QFI \eref{eq:FqChExt} can always
be efficiently evaluated numerically by means of semi-definite programming
\cite{Demkowicz2012} and we show that this is a special case of a more general task of bounding $N$-parallel channels QFI, as explained in \ref{sec:FinNCEasSDP}.
For phase estimation schemes with relevant noise models including: \emph{dephasing},
\emph{depolarization}, \emph{loss} and \emph{spontaneous
emission}, we determine analytically both \eref{eq:maxmin} and \eref{eq:FqChExt}
to verify if the use of extended, entangled input states may improve
estimation at the single channel level. The corresponding unextended/extended
channel QFIs are presented in \Tref{tab:FqCh}, whereas the optimal
purifications yielding the minimum of \eref{eq:FqChExt} can be found
in \ref{sec:ChannelsSpecs} along with the details of the channels considered.
The results justify that extension enhances the precision only for
depolarization and spontaneous emission channels, for which the input states maximally
entangled between $\mathcal{H}_{\textrm{S}}$ and $\mathcal{H}_{\textrm{A}}$
are optimal.
% In other cases, one can use any $\left|\psi_{\textrm{in}}^{\textrm{ext}}\right\rangle $
%such that $\textrm{Tr}_{\mathcal{A}}\!\left\{ \left|\psi_{\textrm{in}}^{\textrm{ext}}\right\rangle \!\left\langle \psi_{\textrm{in}}^{\textrm{ext}}\right|\right\} \!=\!\left|\textrm{+}\right\rangle \!\left\langle \textrm{+}\right|$
%with $\left|\textrm{+}\right\rangle \!=\!\frac{1}{\sqrt{2}}\left(\left|0\right\rangle \!+\!\left|1\right\rangle \right)$
%being also the optimal input for all the above channels when studied
%in their unextended version.

\subsection{RLD-based upper bound on extended channel QFI}
\label{sub:RLD}

In \cite{Hayashi2011} the applicability of the RLD-based bound \eref{eq:FqRLD}
has been addressed in the context of channels. By defining the \emph{Choi-Jamio\l{}kowski}
(C-J) \emph{representation }\cite{Bengtsson2006} of a particular
map $\Lambda_{\varphi}$ as $\Omega_{\Lambda_{\varphi}}\!=\!\Lambda_{\varphi}\otimes\mathcal{I}\left[\left|\mathbb{I}\right\rangle \right]$
with $\left|\mathbb{I}\right\rangle \!=\!\sum_{i=1}^{\dim\mathcal{H}_{\textrm{S}}}\left|i\right\rangle \otimes \left|i\right\rangle $,
it has been proven that the extended channel QFI can be further upper-bounded
by
\begin{equation}
\mathcal{F}\!\left[\Lambda_{\varphi}\otimes\mathcal{I}\right]\le\mathcal{F}^{\textrm{RLD}}\!\left[\Lambda_{\varphi}\otimes\mathcal{I}\right]=\left\Vert \textrm{Tr}_{\mathcal{A}}\!\left\{ \dot{\Omega}_{\Lambda_{\varphi}}\Omega_{\Lambda_{\varphi}}^{-1}\dot{\Omega}_{\Lambda_{\varphi}}\right\} \right\Vert ,\label{eq:FqChExtRLD}
\end{equation}
where $\left\Vert \dots\right\Vert $ is again the operator norm and
$\Omega_{\Lambda_{\varphi}}^{-1}$ is the inverse of $\Omega_{\Lambda_{\varphi}}$
restricted only to the support of the C-J matrix. Most importantly,
the bound \eref{eq:FqChExtRLD} is determined solely by the form of
$\Lambda_{\varphi}$, i.e. its C-J representation, and does not contain
any extra information about the space of input states accepted by
the map. This contrasts the definitions of purification-based unextended/extended
channel QFIs \eref{eq:FqChPurif}/\eref{eq:FqChExt} and facilitates
the analyticity of the results presented in \Tref{tab:FqCh}. On the
other hand, as indicated in \Sref{sub:FqRLD}, both applicability
and tightness of the RLD-based bounds are limited.
The bound
\eref{eq:FqChExtRLD} is valid only when $\dot{\Omega}_{\Lambda_{\varphi}}^{2}$
is fully supported by $\Omega_{\Lambda_{\varphi}}$ \cite{Hayashi2011}.
However, we give an intuitive reason for this restriction by proving
that this condition is equivalent (see \ref{sec:RLDEquivCS}) to the
notion of channel $\Lambda_{\varphi}$ being $\varphi$\emph{-non-extremal},
as introduced in \cite{Demkowicz2012} and also revisited in the
following section. This confirms that the exclusive dependence of
\eref{eq:FqChExtRLD} on $\Omega_{\Lambda_{\varphi}}$ has indeed
a strong geometrical meaning. Moreover, although the RLD-based bounds
depicted in \Tref{tab:FqCh} for the relevant channels seem to be
far above the corresponding channel QFIs---\eref{eq:FqCh} and \eref{eq:FqChExt},
they are of great significance. The bound \eref{eq:FqChExtRLD} is
additive for any maps $\Lambda_{\varphi}^{(1)}$, $\Lambda_{\varphi}^{(2)}$
to which it applies \cite{Hayashi2011},
thus
\begin{eqnarray}
\!\!\!\!\!\!\!\mathcal{F}^{\textrm{RLD}}\!\left[\left(\Lambda_{\varphi}^{(1)}\!\otimes\!\mathcal{I}\right)\!\otimes\!\left(\Lambda_{\varphi}^{(2)}\otimes\mathcal{I}\right)\right] = \mathcal{F}^{\textrm{RLD}}\!\left[\Lambda_{\varphi}^{(1)}\!\otimes\!\mathcal{I}\right]+\mathcal{F}^{\textrm{RLD}}\!\left[\Lambda_{\varphi}^{(2)}\!\otimes\!\mathcal{I}\right]\nonumber \\
\!\!\!\!\!\!\therefore\quad\mathcal{F}\!\left[\left(\Lambda_{\varphi}\!\otimes\!\mathcal{I}\right)^{\otimes N}\right]\le\mathcal{F}^{\textrm{RLD}}\!\!\left[\left(\Lambda_{\varphi}\!\otimes\!\mathcal{I}\right)^{\otimes N}\right]  = N\;\mathcal{F}^{\textrm{RLD}}\!\left[\Lambda_{\varphi}\!\otimes\!\mathcal{I}\right].\label{eq:FqChExtRLDAs}
\end{eqnarray}
Hence, it constrains not only the QFI of a single extended channel,
but also restricts the QFI of $N$ extended channels used in parallel
to scale at most linearly with $N$. Crucially, as the extension can
only improve the precision, \eref{eq:FqChExtRLDAs} is also a valid
upper-bound on the QFI of $N$ uses of an unextended channel, which
asymptotic precision scaling is then limited to a constant factor
improvement over the SQL (see \Sref{sub:SQLAsBounds}).

%%%%%%%%%%%%%%%%%%%%%%%%%%%%%%%%%%%%%%%%%%%%%%%%%%%%%%%%%%%%%%%%%%%%%%%%%%%%%%%%%%%%%%%%%%%5

\section{Estimation of $N$ independent quantum channels \label{sec:QFIChN}}
\label{sec:EstNChannels}

\begin{table}[!t]
\begin{tabular}{|>{\centering}m{2.5cm}|>{\centering}m{1cm}|>{\centering}m{1.6cm}||>{\centering}m{1.8cm}|>{\centering}m{1.7cm}|>{\centering}m{2.25cm}|>{\centering}m{1.75cm}|>{\centering}m{1.6cm}}
\cline{1-7}
\textbf{Noise model} & $\!\mathcal{F}\!\left[\Lambda_{\varphi}\right]$ & $\!\mathcal{F}\!\left[\Lambda_{\varphi}\!\otimes\!\mathcal{I}\right]$ & $\!\mathcal{F}_{\textrm{as}}^{\textrm{CE}}${ \footnotesize in \cite{Demkowicz2012}}&
 $\mathcal{F}_{\textrm{as}}^{\textrm{QS}}$ & $\!\mathcal{F}^{\textrm{RLD}}\!\left[\Lambda_{\varphi}\!\otimes\!\mathcal{I}\right]$&
$\!\mathcal{F}_{\textrm{as}}^{\textrm{CS}}${ \footnotesize in \cite{Demkowicz2012}} & \tabularnewline
\cline{1-7}
\emph{Dephasing} & $\eta^{2}$ & $\eta^{2}$ & $\frac{\eta^{2}}{1-\eta^{2}}$ & $\frac{\eta^{2}}{1-\eta^{2}}$ & $\frac{\eta^{2}}{1-\eta^{2}}$ & $\frac{\eta^{2}}{1-\eta^{2}}$ & ~\\
\tabularnewline
\cline{1-7}
\emph{\negthinspace{}Depolarization} & $\eta^{2}$ & $\frac{2\eta^{2}}{1+\eta}$ & $\!\frac{2\eta^{2}}{\left(1-\eta\right)\left(1+2\eta\right)}$ & $\!\frac{2\eta^{2}}{\left(1-\eta\right)\left(1+2\eta\right)}$ & $\frac{2\eta^{2}(1+\eta)}{\left(1-\eta\right)\left(1+3\eta\right)}$ & $\!\frac{4\eta^{2}}{\left(1-\eta\right)\left(1+3\eta\right)}$ & ~\\
\tabularnewline
\cline{1-7}
\emph{Loss} & $\eta$ & $\eta$ & $\frac{\eta}{1-\eta}$ & $\frac{\eta}{1-\eta}$ & n.a. & n.a. & ~\\ \tabularnewline
\cline{1-7}
\emph{Spontaneous emission} & $\eta$ & $\frac{4\eta}{\left(1+\sqrt{\eta}\right)^{2}}$ & $\frac{4\eta}{1-\eta}$ & n.a. & n.a. & n.a. & \tabularnewline
\cline{1-7}
\end{tabular}\caption{\label{tab:FqCh} \textbf{Channel phase estimation sensitivity quantified via QFIs and their asymptotic bounds.}
 The noise models of metrological relevance discussed in the paper are
listed in the first column. Decoherence strength increases with a decrease of the $\eta$ parameter
 $(0\!\le\!\eta\!<\!1)$ (see \ref{sec:ChannelsSpecs} for details).
From left to right---single channel QFI, extended channel QFI,  upper bounds on asymptotic channel QFI \eref{eq:FqAs} in ascending order:
CE bound (see \Sref{sub:CEmethod}), QS bound (see \Sref{sub:QS}), RLD-based bound
(see \Sref{sub:RLD}), CS bound (see \Sref{sub:CS}). \hfill[n.a.---not~available]}
\end{table}
%For each channel, the value of the asymptotic channel QFI, \eref{eq:FqAs},
%determining the precision in the $N\!\rightarrow\!\infty$ limit is
%confined to lie in the region constrained by the vertical double-lines.

In order to describe general metrological schemes depicted in \Fref{fig:qm_scheme}\textbf{a},
%incorporating uncorrelated-noise effects
we model the evolution of all particles within the probe as $N$
identical, independent channels acting on a possibly entangled, pure
input state of the whole probe $\left|\psi_{\textrm{in}}^{N}\right\rangle $.
The final output state of the probe then reads $\rho_{\varphi}^{N}\!=\!\Lambda_{\varphi}^{\otimes N}\!\left[\left|\psi_{\textrm{in}}^{N}\right\rangle \right]$
yielding the $N$-\emph{channel QFI}:
\begin{equation}
\mathcal{F}\!\left[\Lambda_{\varphi}^{\otimes N}\right]=\max_{\psi_{\textrm{in}}^{N}}\, F_{\textrm{Q}}\!\left[\,\Lambda_{\varphi}^{\otimes N}\!\left[\left|\psi_{\textrm{in}}^{N}\right\rangle \right]\right],\label{eq:FqNCh}
\end{equation}
which linear or quadratic dependence on $N$ dictates respectively
the SQL or HL scaling of precision. For example, when considering
classical schemes that employ unentangled probes, $\left|\psi_{\textrm{in}}^{N}\right\rangle \!=\!\otimes_{n=1}^{N}\left|\psi_{\textrm{in}}^{1}\right\rangle $,
we are effectively dealing with $N$ independent subsystems, so that
$\mathcal{F}\!\left[\Lambda_{\varphi}^{\otimes N}\right]\!=\! N\,\mathcal{F}\!\left[\Lambda_{\varphi}\right]$
and the uncertainty of the estimate $\tilde{\varphi}$ decreases as
$1/\sqrt{N}$.

\subsection{Standard quantum limit (SQL)-like bounds on precision in the asymptotic $N$ limit}
\label{sub:SQLAsBounds}

To investigate channels that incorporate the uncorrelated noise restricting
the asymptotic precision scaling to SQL, we define the \emph{asymptotic
channel QFI }as
\begin{equation}
\mathcal{F}_{\textrm{as}}\!\left[\Lambda_{\varphi}\right]=\lim_{N\rightarrow\infty}\frac{\mathcal{F}\!\left[\Lambda_{\varphi}^{\otimes N}\right]}{N}.\label{eq:FqAs}
\end{equation}
Thus, for such SQL-bound channels, \eref{eq:FqAs} is finite and $\mathcal{F}_{\textrm{as}}\!\left[\Lambda_{\varphi}\right]\!\ge\!\mathcal{F}\!\left[\Lambda_{\varphi}\right]$
with equality indicating the optimality of classical estimation schemes.
Hence, \eref{eq:FqAs} quantifies the \emph{maximal quantum precision enhancement}
%over classical strategies
that reads
\begin{equation}
\chi\!\left[\Lambda_{\varphi}\right]=\lim_{N\rightarrow\infty}\frac{\Delta\tilde{\varphi}_{\textrm{cl}}}{\Delta\tilde{\varphi}_{\textrm{Q}}}=\sqrt{\frac{\mathcal{F}_{\textrm{as}}\!\left[\Lambda_{\varphi}\right]}{\mathcal{F}\!\left[\Lambda_{\varphi}\right]}}\ge1\label{eq:qEnh}.
\end{equation}
%and is locally achievable as $N\!\rightarrow\!\infty$ (possibly requiring
%$k\!\rightarrow\!\infty$ to saturate \eref{eq:clCRB} and \eref{eq:qCRB}).
However, as the computation of \eref{eq:FqAs} is generally infeasible
owing to the complexity of QFI rising exponentially with $N$, one
normally needs to construct an upper limit on the $N$-channel QFI \eref{eq:FqNCh}, $\mathcal{F}^{\textrm{bound}}[\Lambda_\varphi^{\otimes N}]$,
 from which the asymptotic form, $\mathcal{F}^{\textrm{bound}}_{\textrm{as}}$,
may be deduced using \eref{eq:FqAs} that upper-bounds both the $N$-channel QFI and the maximal quantum precision enhancement:
%Then, for a given SQL-bound channel, one can formulate upper bounds
%on \eref{eq:FqNCh} and \eref{eq:qEnh} respectively:
\begin{equation}
\mathcal{F}\!\left[\Lambda_{\varphi}^{\otimes N}\right]\le N\,
\mathcal{F}_{\textrm{as}}^{\textrm{bound}},\qquad\qquad\chi\!\left[\Lambda_{\varphi}\right]\le\sqrt{\frac{\mathcal{F}_{\textrm{as}}^{\textrm{bound}}\!
\left[\Lambda_{\varphi}\right]}{\mathcal{F}\!\left[\Lambda_{\varphi}\right]}}.\label{eq:FqNChAsBound}
\end{equation}
%of which the second one may be attained by some estimation protocol
%in the asymptotic $N$ limit, only if the conjectured asymptotic bound
%is tight, i.e. $\mathcal{F}_{\textrm{as}}^{\textrm{bound}}\!=\!\mathcal{F}_{\textrm{as}}$.

Methods of constructing $\mathcal{F}_{\textrm{as}}^{\textrm{bound}}$
were proposed in \cite{Demkowicz2012} basing on the concepts of
channel \emph{classical simulation} (CS) and \emph{channel extension}
(CE). As mentioned already, the CS method applies only to $\varphi$\emph{-non-extremal}
channels, for which also the RLD-based bound \eref{eq:FqChExtRLDAs}
provides a valid example of $\mathcal{F}_{\textrm{as}}^{\textrm{bound}}$. Yet, the notion of CS
may be generalized to the idea of channel \emph{quantum simulation} (QS) introduced in
\cite{Matsumoto2010}, in order to obtain an asymptotic bound applicable
to a wider class of quantum maps. All these four approaches ($\mathcal{F}_{\textrm{as}}^{\textrm{bound}}\!=\!\mathcal{F}_{\textrm{as}}^{\textrm{CS}},\mathcal{F}_{\textrm{as}}^{\textrm{CE}},\mathcal{F}^{\textrm{RLD}}\!\left[\Lambda_{\varphi}\!\otimes\!\mathcal{I}\right],\mathcal{F}_{\textrm{as}}^{\textrm{QS}}$
respectively) are presented in \Tref{tab:FqCh} on the right hand
side of the double-line for the relevant channels. As the lossy
and spontaneous emission interferometry cases are examples of $\varphi$-extremal\emph{
}maps, they do not allow for finite $\mathcal{F}_{\textrm{as}}^{\textrm{CS}}$
and $\mathcal{F}^{\textrm{RLD}}\!\left[\Lambda_{\varphi}\!\otimes\!\mathcal{I}\right]$
to be constructed. In the case of depolarization channel, which is
full-rank \cite{Bengtsson2006} and hence $\varphi$-non-extremal,
despite the lack of a simple geometric interpretation of its value,
$\mathcal{F}^{\textrm{RLD}}\!\left[\Lambda_{\varphi}\!\otimes\!\mathcal{I}\right]$
proves to be tighter than $\mathcal{F}_{\textrm{as}}^{\textrm{CS}}$.
The more general QS approach not only applies also to the ($\varphi$-extremal)
lossy interferometry case, but also provides\emph{ }as accurate
bounds as the CE method. Nevertheless, as the CE approach is proven
to provide at least as tight bounds for the broadest class of quantum
channels containing ones to which the other methods apply, we use
it to quantify the maximal quantum precision enhancements \eref{eq:qEnh} for
the channels considered (see \Tref{tab:qEnh}). Below, we describe
alternately the CS, QS and CE methods---ordered according to their power
and generality.

\subsubsection{Classical simulation (CS) method}
\label{sub:CS}
~\\
As introduced in \cite{Matsumoto2010} and depicted in \Fref{fig:qm_scheme}\textbf{b},
a channel admits a CS of itself, if
for any $\varrho$ it is expressible in the form
\begin{equation}
\Lambda_{\varphi}\!\left[\varrho\right]=\Phi\!\left[\varrho\otimes p_{\varphi}\right]=\sum_{i}p_{\varphi,i}\,\Pi_{i}\!\left[\varrho\right],\label{eq:CS}
\end{equation}
where $p_{\varphi}\!=\!\sum_{i}p_{\varphi,i}\left|e_{i}\right\rangle \!\left\langle e_{i}\right|$
is a classical, diagonal density matrix in some basis and $\Phi$
is a $\varphi$-independent CPTP map acting on a larger input space.
By defining $\Pi_{i}\!\left[\varrho\right]\!=\!\Phi\!\left[\varrho\!\otimes\!\left|e_{i}\right\rangle \!\left\langle e_{i}\right|\right]$
one obtains the second equality in \eref{eq:CS}, so that it becomes
evident that the estimated $\varphi$ parametrizes only the mixing
probabilities of some $\varphi$-independent quantum maps. Then, the
$N$-channel QFI \eref{eq:FqNCh} can be simply upper-bounded via
$\mathcal{F}\!\left[\Lambda_{\varphi}^{\otimes N}\right]\!\le\! N\, F_{\textrm{cl}}\!\left[p_{\varphi,i}\right]$,
where $F_{\textrm{cl}}\!\left[p_{\varphi,i}\right]$ is the discrete
version of classical FI in \eref{eq:clCRB} and plays the role of
$\mathcal{F}_{\textrm{as}}^{\textrm{bound}}$ in \eref{eq:FqNChAsBound}
\cite{Matsumoto2010, Demkowicz2012}. Moreover, as QFI is a local quantity, in order to construct a CS-based $\mathcal{F}_{\textrm{as}}^{\textrm{bound}}$
valid for small deviations $\delta\varphi$ around a given $\varphi$,
the classical simulation must be feasible only locally, i.e. $\Lambda_{\varphi}\!\left[\varrho\right]\!=\!\sum_{i}p_{\varphi,i}\,\Pi_{i}\!\left[\varrho\right]+O(\delta\varphi^{2})$.
Therefore, as proven in \cite{Matsumoto2010}, if the C-J representation
$\Omega_{\Lambda_{\varphi}}$ of a channel $\Lambda_{\varphi}$ allows
for parameters $\epsilon_{\pm}\!>\!0$ such that the matrices $\Omega_{\Pi_{\pm}}\!=\!\Omega_{\Lambda_{\varphi}}\pm\,\epsilon_{\pm}\dot{\Omega}_{\Lambda_{\varphi}}$
are positive semi-definite at a given $\varphi$, the channel is $\varphi$\emph{-non-extremal}
there and the necessary $p_{\varphi,i}$ can always be found. This
is because one can construct $\Omega_{\tilde{\Lambda}_{\varphi}}\!=\! p_{\varphi,+}\Omega_{\Pi_{+}}+p_{\varphi,-}\Omega_{\Pi_{-}}$
that up to $O(\delta\varphi^{2})$ coincides with $\Omega_{\Lambda_{\varphi}}$
by choosing $p_{\varphi,\pm}$ such that $\Omega_{\tilde{\Lambda}_{\varphi}}\!\!=\!\Omega_{\Lambda_{\varphi}}$
and $\dot{\Omega}_{\tilde{\Lambda}_{\varphi}}\!\!=\!\dot{\Omega}_{\Lambda_{\varphi}}$.
Hence, $\Lambda_{\varphi}\!\left[\varrho\right]\!=\!\tilde{\Lambda}_{\varphi}\!\left[\varrho\right]+O(\delta\varphi^{2})$
with $\tilde{\Lambda}_{\varphi}\!\left[\varrho\right]\!=\! p_{\varphi,+}\Pi_{+}+p_{\varphi,-}\Pi_{-}$,
so that $F_{\textrm{cl}}\!\left[p_{\varphi,\pm}\right]\!=\!1/(\epsilon_{+}\epsilon_{-})$
is a legal example of the required finite bound valid at $\varphi$. Furthermore,
in \cite{Demkowicz2012}, it has been proven that for channels of
the form $\Lambda_{\varphi}\!\left[\varrho\right]\!=\!\Lambda\!\left[\,\mathcal{U}_{\varphi}\!\left[\varrho\right]\right]$
this two-point construction is always optimal at any $\varphi$ when
maximal possible $\epsilon_{\pm}$ are chosen
\footnote{Yet, it may prove optimal for channels of other type, as shown for
transversal dephasing in \cite{Chaves2012}.}.
Geometrically, imagining the convex set of all CPTP maps in their
C-J representation that share input and output spaces of $\Lambda_{\varphi}$,
the channels $\Omega_{\Pi_{\pm}}$ lie at the intersection points
of the tangent generated by $\dot{\Omega}_{\Lambda_{\varphi}}$ at
$\Omega_{\Lambda_{\varphi}}$ and the boundary of the set. The $\mathcal{F}_{\textrm{as}}^{\textrm{bound}}$
of \eref{eq:FqNChAsBound}, which we refer to as the \emph{asymptotic
CS bound}---$\mathcal{F}_{\textrm{as}}^{\textrm{CS}}\!=\!1/(\epsilon_{+}^{\textrm{max}}\epsilon_{-}^{\textrm{max}})$,
is dictated then by the ``distances'' $\epsilon_{\pm}^{\textrm{max}}$
of the channel to the boundary measured along this tangent. Although
the CS approach provides weaker bounds than the CE method \cite{Demkowicz2012},
it gives an intuitive geometric explanation of the inevitable asymptotic
SQL-like scaling of all $\varphi$-non-extremal maps. These naturally
include the full-rank channels \cite{Bengtsson2006}, which lie inside the set of CPTP maps
away from its boundary.

\subsubsection{Quantum simulation (QS) method}
\label{sub:QS}
~\\
In \cite{Matsumoto2010}, a natural generalization of the channel
CS has been proposed which is schematically presented in \Fref{fig:qm_scheme}\textbf{b}.
This, so called, QS of a channel corresponds
to expressing its action in a form similar to \eref{eq:CS} that reads
\begin{equation}
\Lambda_{\varphi}\!\left[\varrho\right]=\Phi\!\left[\varrho\otimes\sigma_{\varphi}\right]=
\textrm{Tr}_{\textrm{E}_{\Phi}\textrm{E}_{\sigma}}\!\!\left\{ U\left(\varrho\otimes
\left|\psi_{\varphi}\right\rangle_{\textrm{E}_{\Phi}\textrm{E}_{\sigma}} \!\!\left\langle \psi_{\varphi}\right|\right)U^{\dagger}\right\} \!,\label{eq:QS}
\end{equation}
where now $\sigma_{\varphi}$ is a quantum, non-diagonal, $\varphi$-dependent
density matrix, and the purified form on the right hand side involves both channel $\Phi$ environment space $\textrm{E}_\Phi$
as well as $\sigma_{\varphi}$ purification space $\textrm{E}_\sigma$, such that
$\sigma_{\varphi}\!=\!\textrm{Tr}_{\textrm{E}_{\sigma}}\!\left\{ \left|\psi_{\varphi}\right\rangle \!\left\langle \psi_{\varphi}\right|\right\} $.
Note that the notion of \emph{quantum simulability}
is equivalent to the \emph{channel} \emph{programmability} concept
introduced in \cite{Ji2008}. Following the same argumentation as
in \cite{Demkowicz2012} for CS, the $N$-channel QFI \eref{eq:FqNCh}
of a \emph{quantum simulable channel}---one admitting a QS of the form \eref{eq:QS} with finite
$F_{\textrm{Q}}\!\left[\sigma_{\varphi}\right]$---can be linearly bounded
as $\mathcal{F}\!\left[\Lambda_{\varphi}^{\otimes N}\right]\!\le\! N\, F_{\textrm{Q}}\!\left[\sigma_{\varphi}\right]$,
and therefore the asymptotic bound reads $\mathcal{F}_{\textrm{as}}^{\textrm{bound}}\!=\!F_{\textrm{Q}}[\sigma_\varphi]$.
Similarly to CS, a channel may admit many decompositions \eref{eq:QS}
and the optimal one must yield the lowest $F_{\textrm{Q}}\!\left[\sigma_{\varphi}\right]$.
Therefore, without loss of generality, in the search for the optimal QS,
we may take $U$ in \eref{eq:QS} to act on the full purified system, i.e. also in $\textrm{E}_{\Phi}$
and $\textrm{E}_{\sigma}$ spaces. This enlarges the set of all possible
QSs beyond the original ones $U\!=\! U^{\textrm{S}\textrm{E}_{\Phi}}\!\otimes\!\mathbb{I}^{\textrm{E}_{\sigma}}$
and yields $\mathcal{F}_{\textrm{as}}^{\textrm{bound}}\!=\! F_{\textrm{Q}}\!\left[\left|\psi_{\varphi}\right\rangle \right]$,
which via \eref{eq:FqPurifEscher} cannot be smaller than $F_{\textrm{Q}}\!\left[\sigma_{\varphi}\right]$.
In fact, \eref{eq:FqPurifEscher} assures that for any QS employing
$\sigma_{\varphi}$, there exists an ``enlarged'' decomposition
\eref{eq:QS} leading to the same $\mathcal{F}_{\textrm{as}}^{\textrm{bound}}$
with $\ket{\psi_{\varphi}}$ being the minimal purification in \eref{eq:FqPurifEscher}.
In conclusion, we may seek the optimal QS by analysing all possible
decompositions of the form \eref{eq:QS} that, owing to the locality
of the QFI, must be feasible only for small
deviations $\delta\varphi$ from a given $\varphi$, so that $\Lambda_{\varphi}\!\left[\varrho\right]\!=\!\textrm{Tr}_{\textrm{E}_{\Phi}\textrm{E}_{\sigma}}\!\!\left\{ U\left(\varrho\!\otimes\!\left|\psi_{\varphi}\right\rangle \!\left\langle \psi_{\varphi}\right|\right)U^{\dagger}\right\} +O(\delta\varphi^{2})$.
In \ref{sec:QSasCE}, we prove that, in order for the QS \eref{eq:QS}
to be possible locally at $\varphi$ and lead to a finite asymptotic bound, $\Lambda_{\varphi}$ of rank
$r$ must admit Kraus operators $\left\{ K_{i}(\varphi)\right\} _{i=1}^{r}$
that satisfy the two conditions:
\begin{equation}
\!\!\!\!\!\!\!\!\!\!\!\!\!\!\!\!\!\!\!\!\textrm{i}\sum_{i=1}^{r}\dot{K}_{i}(\varphi)^{\dagger}K_{i}(\varphi)=0\qquad\textrm{and}\qquad\sum_{i=1}^{r}\dot{K}_{i}(\varphi)^{\dagger}\dot{K}_{i}(\varphi)=\frac{1}{4}F_{\textrm{Q}}\!\left[\left|\psi_{\varphi}\right\rangle \right]\,\mathbb{I}.\label{eq:QScontrs}
\end{equation}
Hence, by optimizing over all locally equivalent Kraus representations
of $\Lambda_{\varphi}$---the ones related to one another by rotations
\eref{eq:KrausReps} generated by any Hermitian $h$---that
satisfy constraints \eref{eq:QScontrs}, we may determine the
asymptotic bound given by the optimal local QS, which we refer to
as the \emph{asymptotic QS bound}---$\mathcal{F}_{\textrm{as}}^{\textrm{QS}}$,
as follows
\begin{equation}
\mathcal{F}_{\textrm{as}}^{\textrm{QS}}=\min_{h}\,\lambda\quad\textrm{ s.t. \;}
\;\alpha_{\tilde{K}}=\frac{\lambda}{4}\,\mathbb{I},\;
\beta_{\tilde{K}}=0,\label{eq:F_QS}
\end{equation}
where $\alpha_{\tilde{K}}\!=\!\sum_{i=1}^{r}\dot{\tilde{K}}_{i}(\varphi)^{\dagger}\dot{\tilde{K}}_{i}(\varphi)$,
$\beta_{\tilde{K}}\!=\!\textrm{i}\sum_{i=1}^{r}\dot{\tilde{K}}_{i}(\varphi)^{\dagger}\tilde{K}_{i}(\varphi)$
and $\lambda$ has the interpretation of $\mathcal{F}_{\textrm{as}}^{\textrm{bound}}\!=\!F_{\textrm{Q}}\!\left[\left|\psi_{\varphi}\right\rangle \right]$
for a local QS of the form \eref{eq:QS}. Before revisiting the
CE method explicitly below, we should note that \eref{eq:F_QS} resembles
exactly the asymptotic CE bound of \cite{Demkowicz2012} with an
extra constraint forcing the operator $\alpha_{\tilde{K}}$ to be
proportional to identity. This proves that indeed the QS method can
never outperform the CE approach.

Investigating the relevant quantum maps considered in \Tref{tab:FqCh},
the QS method must naturally apply to dephasing and depolarization
channels. These are $\varphi$-non-extremal
maps, hence their locally constructible CSs \eref{eq:CS} serve as
examples of the more general QSs \eref{eq:QS}. Consistently, the
Kraus representations utilized in \cite{Demkowicz2012} to obtain
the asymptotic CE bounds for these two channels fulfil the $\alpha_{\tilde{K}}\!\propto\!\mathbb{I}$
constraint. Thus, QS is not only feasible in their case but also its
asymptotic bound coincides with the one of the superior CE method.
Significantly, also in the case of the lossy interferometry
the optimal Kraus operators used in \cite{Demkowicz2012} to minimize
the asymptotic CE bound satisfy the extra QS's constraint. This fact
indicates that also for $\varphi$-extremal channels QS may prove
to be as good as CE. However, in the case of spontaneous emission, the QS method seizes to work, as the $\beta_{\tilde{K}}\!=\!0$
condition fixes $\alpha_{\tilde{K}}$ to be disproportional to identity
\cite{Demkowicz2012}.

\subsubsection{Channel extension (CE) method}
\label{sub:CEmethod}

\begin{table}[!t]
\begin{tabular}{|>{\centering}m{2.5cm}|>{\centering}m{2.5cm}|>{\centering}m{3cm}|>{\centering}m{2.5cm}|>{\centering}m{2.5cm}|>{\centering}m{2.5cm}}
\cline{1-5}
\textbf{Noise model} & \emph{Dephasing} & \emph{Depolarization} & \emph{Loss } & \emph{Spontaneous emission} & \tabularnewline
\cline{1-5}
$\chi\!\left[\Lambda_{\varphi}\right]$ & $\!=\sqrt{\frac{1}{1-\eta^{2}}}$ & $\le\sqrt{\frac{2}{(1-\eta)(1+2\eta)}}$ & $\!=\sqrt{\frac{1}{1-\eta}}$ & $\!\le\sqrt{\frac{4}{1-\eta}}$ & ~\\
\tabularnewline
\cline{1-5}
$\chi\!\left[\Lambda_{\varphi}\otimes\mathcal{I}\right]$ & $\!=\sqrt{\frac{1}{1-\eta^{2}}}$ & $=\sqrt{\frac{1+\eta}{(1-\eta)(1+2\eta)}}$ & $\!=\sqrt{\frac{1}{1-\eta}}$ & $\!=\sqrt{\frac{1+\sqrt{\eta}}{1-\sqrt{\eta}}}$ & ~\\
\tabularnewline
\cline{1-5}
\end{tabular}\caption{\label{tab:qEnh} \textbf{Quantum phase estimation precision enhancement from the
CE method.} For all noise models specified in \ref{sec:ChannelsSpecs}, the asymptotic
CE bounds on the maximal quantum precision enhancement factors, $\chi\!\left[\bullet\right]\!=\!\sqrt{\mathcal{F}_{\textrm{as}}\!\left[\bullet\right]/\mathcal{F}\!\left[\bullet\right]}$,
are presented. For a general quantum map $\Lambda_{\varphi}$, the
CE bound only upper-limits $\chi\!\left[\Lambda_{\varphi}\right]$
as $\mathcal{F}_{\textrm{as}}\!\left[\Lambda_{\varphi}\right]\!\le\!\mathcal{F}_{\textrm{as}}^{\textrm{CE}}$.
Yet, for dephasing and lossy interferometry, as indicated
by ``$=$'', the corresponding values of $\chi\!\left[\Lambda_{\varphi}\right]$
have been shown to be attainable \cite{Orgikh2001,Caves1981}.
For an extended channel, $\chi\!\left[\Lambda_{\varphi}\otimes\mathcal{I}\right]$
is determined by the CE bound as $\mathcal{F}_{\textrm{as}}\!\left[\Lambda_{\varphi}\otimes\mathcal{I}\right]\!=\!\mathcal{F}_{\textrm{as}}^{\textrm{CE}}$.}
\end{table}

~\\
The CE method of \cite{Demkowicz2012}
not only applies to the widest class of quantum maps containing all
$\varphi$-non-extremal ones, but also provides more stringent bounds
than the CS, RLD and QS equivalents, as respectively proven in \cite{Demkowicz2012},
\ref{sec:RLDasCE} and above. In this method, see \Fref{fig:qm_scheme}\textbf{b},
each single channel is extended by an auxiliary ancilla as introduced
in \Sref{sub:QFIChExt}. In \cite{Fujiwara2008}, it has been proven
that one can then effectively bound the $N$-channel QFI \eref{eq:FqNCh}
via the $N$\emph{-extended-channel QFI}, so that
\begin{equation}
\!\!\!\!\!\!\!\!\!\!\!\!\!\!\!\!\!\!\!\!\!\mathcal{F}\!\left[\Lambda_{\varphi}^{\otimes N}\right]\le\mathcal{F}\!\left[\left(\Lambda_{\varphi}\otimes\mathcal{I}\right)^{\otimes N}\right]\le4\,\min_{h}\left\{ N\,\left\Vert \alpha_{\tilde{K}}\right\Vert +N(N-1)\left\Vert \beta_{\tilde{K}}\right\Vert ^{2}\right\} ,\label{eq:FujiBound}
\end{equation}
where again $\alpha_{\tilde{K}}\!=\!\sum_{i=1}^{r}\dot{\tilde{K}}_{i}(\varphi)^{\dagger}\dot{\tilde{K}}_{i}(\varphi)$,
$\beta_{\tilde{K}}\!=\!\textrm{i}\sum_{i=1}^{r}\dot{\tilde{K}}_{i}(\varphi)^{\dagger}\tilde{K}_{i}(\varphi)$
and $h$ is the generator of local Kraus representation rotations
\eref{eq:KrausReps}. Crucially, if there exists a Kraus representation
for which the second term in \eref{eq:FujiBound} vanishes, $\mathcal{F}\!\left[\Lambda_{\varphi}^{\otimes N}\right]$
must asymptotically scale linearly in $N$. This requirement corresponds
to the constraint $\beta_{\tilde{K}}\!=\!0$ already accounted in
the QS method, which for any linearly independent Kraus operators
$\{K_{i}\}_{i=1}^{r}$ is equivalent to the existence of $h$
such that \cite{Fujiwara2008}
\begin{equation}
\sum_{i,j=1}^{r}h_{ij}\, K_{i}^{\dagger}K_{j}=\textrm{i}\sum_{i=1}^{r}\dot{K}_{i}(\varphi)^{\dagger}K_{i}(\varphi).\label{eq:FIC}
\end{equation}
What is more, for any channel that admits an $h$ fulfilling
\eref{eq:FIC}, one can show basing on the results of \cite{Fujiwara2008}
that the second inequality in \eref{eq:FujiBound} is saturated in
the $N\!\rightarrow\!\infty$ limit, so that the \emph{asymptotic
extended channel QFI }then reads
\begin{equation}
\!\!\!\!\!\!\!\!\!\!\!\!\!\!\!\!\!\!\!\!\!\mathcal{F}_{\textrm{as}}\!\left[\Lambda_{\varphi}\otimes\mathcal{I}\right]=
\lim_{N\rightarrow\infty}\!\frac{\mathcal{F}\!\left[\left(\Lambda_{\varphi}\otimes\mathcal{I}\right)^{\otimes N}\right]}{N}=
4\min_{\underset{\beta_{\tilde{K}}=0}{h}}\,\left\Vert \sum_{i=1}^{r}\dot{\tilde{K}}_{i}(\varphi)^{\dagger}\dot{\tilde{K}}_{i}(\varphi)\right\Vert .\label{eq:FqCE}
\end{equation}
Importantly, \eref{eq:FqCE} becomes the required asymptotic bound
$\mathcal{F}_{\textrm{as}}^{\textrm{bound}}$ of \eref{eq:FqNChAsBound},
which we refer to as the \emph{asymptotic CE bound}---$\mathcal{F}_{\textrm{as}}^{\textrm{CE}}$.
We explicitly wrote the full form of \eref{eq:FqCE} to emphasize
its similarity to the extended \emph{single} channel QFI \eref{eq:FqChExt}. The
essential difference in \eref{eq:FqCE} is the constraint \eref{eq:FIC}
yielding $\mathcal{F}_{\textrm{as}}\!\left[\Lambda_{\varphi}\otimes\mathcal{I}\right]\!\ge\!\mathcal{F}\!\left[\Lambda_{\varphi}\otimes\mathcal{I}\right]$
and dictating the maximal quantum precision enhancement for
an extended channel:
\begin{equation}
\chi\!\left[\Lambda_{\varphi}\otimes\mathcal{I}\right]=\lim_{N\rightarrow\infty}\frac{\Delta\tilde{\varphi}_{\textrm{cl}}^{\textrm{ext}}}{\Delta\tilde{\varphi}_{\textrm{Q}}^{\textrm{ext}}}=\sqrt{\frac{\mathcal{F}_{\textrm{as}}\!\left[\Lambda_{\varphi}\otimes\mathcal{I}\right]}{\mathcal{F}\!\left[\Lambda_{\varphi}\otimes\mathcal{I}\right]}}\ge1.\label{eq:qEnhExt}
\end{equation}
Similarly to \eref{eq:FqChExt}, \eref{eq:FqCE} is computable by
means of semi-definite programming \cite{Demkowicz2012}, so that
one can efficiently determine both \eref{eq:FqNChAsBound} and \eref{eq:qEnhExt}.
The CE-based bounds on $\chi\!\left[\Lambda_{\varphi}\right]$
and the exact values of $\chi\!\left[\Lambda_{\varphi}\otimes\mathcal{I}\right]$
for the relevant noise models are presented in \Tref{tab:qEnh}. Although generally the CE method
only upper-limits the maximal quantum precision enhancement \eref{eq:qEnh},
it has been proven to quantify $\chi\!\left[\Lambda_{\varphi}\right]$
exactly in the case of dephasing \cite{Orgikh2001} and lossy interferometer
channels \cite{Caves1981}. This has been achieved by showing that
input states utilizing spin- and light- squeezing respectively yield
a quantum enhancement that asymptotically attains the corresponding
CE-based bounds presented in \Tref{tab:qEnh}. On the other hand,
as indicated in \Tref{tab:FqCh}, these channels are also examples
of ones for which the extension does not improve the precision at
the single channel level, so that $\chi\!\left[\Lambda_{\varphi}\right]\!=\!\chi\!\left[\Lambda_{\varphi}\otimes\mathcal{I}\right]$
in \Tref{tab:qEnh}. The question---when the lack of precision improvement due to extension at the single channel level
translates to the asymptotic regime, i.e $\mathcal{F}\!\left[\Lambda_{\varphi}\right]\!=\!\mathcal{F}\!\left[\Lambda_{\varphi}\otimes\mathcal{I}\right]$
$\underset{?}{\Longleftrightarrow}$ $\mathcal{F}_{\textrm{as}}\!\left[\Lambda_{\varphi}\right]\!=\!\mathcal{F}_{\textrm{as}}\!
\left[\Lambda_{\varphi}\otimes\mathcal{I}\right]$, we leave open for future research.

\subsection{Finite-$N$ channel extension (CE) method}
\label{sub:UnitWithDeco}

In \Sref{sub:SQLAsBounds}, we have presented the CE method as the
most effective one out of all discussed that provides the tightest
upper limits on the maximal possible asymptotic quantum precision enhancement.
 On the other hand, in the case
of experiments such as optical interferometry with moderate numbers
of photons \cite{Mitchell2004,Nagata2007}, the asymptotic CE bounds,
despite still being valid, are far too weak to be useful. For very low values of $N$, the precision can be quantified
numerically, for instance by brute-force type methods computing explicitly
the QFI. However, in the intermediate $N$ regime---being beyond
the reach of computational power, yet with $N$ too low for the asymptotic
methods to be effective---more accurate bounds should play an important
role.

We propose the finite-N CE method which, despite
being based on the properties of a single channel, still provides
bounds on precision that are relevant in the intermediate $N$ regime.\emph{
}We utilize the upper-limit \eref{eq:FujiBound} on the $N$-extended-channel
QFI and construct the \emph{finite-N CE bound}, $\mathcal{F}_{N}^{\textrm{CE}}$,
that reads
\begin{equation}
\!\!\!\!\!\!\!\!\!\!\!\!\!\frac{\mathcal{F}\!\left[\left(\Lambda_{\varphi}\otimes\mathcal{I}\right)^{\otimes N}\right]}{N}\;
\le\quad\mathcal{F}_{N}^{\textrm{CE}}=4\,\min_{h}\left\{ \left\Vert \alpha_{\tilde{K}}\right\Vert +(N-1)\left\Vert \beta_{\tilde{K}}\right\Vert ^{2}\right\} \!.\label{eq:FqCEN}
\end{equation}
Following the suggestion of \cite{Demkowicz2012}, in contrast to
the asymptotic CE bound $\mathcal{F}_{\textrm{as}}^{\textrm{CE}}$
defined in \eref{eq:FqCE}, we do not impose the SQL-bounding condition
$\beta_{\tilde{K}}\!=\!0$ \eref{eq:FIC}, but we seek at each $N$
for the minimal Kraus representation that is generated by some optimal
$h\!=\!h_{\textrm{opt}}(N)$ being now not only
channel but also $N$-dependent. Still, as shown in \ref{sec:FinNCEasSDP},
$\mathcal{F}_{N}^{\textrm{CE}}$ can always be efficiently evaluated
numerically by recasting the minimization over $h$ in \eref{eq:FqCEN}
into a semi-definite programming task. Moreover, as the finite-$N$ CE
bound varies smoothly between $N\!=\!1$ and $N\!=\!\infty$, at which
it is tight, it provides more accurate bounds than its asymptotic
version.
%, especially for quantum maps that do not experience precision gain
%due to extension at the single channel level --- with $\mathcal{F}\!\left[\Lambda_{\varphi}\right]\!=\!\mathcal{F}\!\left[\Lambda_{\varphi}\otimes\mathcal{I}\right]$
%(e.g. dephasing and lossy interferometer channels)
 For $N\!=\!1$, $\mathcal{F}_{N}^{\textrm{CE}}$ coincides with
the extended channel QFI \eref{eq:FqChExt}---$\mathcal{F}_{N=1}^{\textrm{CE}}\!=\!\mathcal{F}\!\left[\Lambda_{\varphi}\otimes\mathcal{I}\right]$,
whereas in the $N\!\rightarrow\!\infty$ limit it attains the asymptotic
CE bound \eref{eq:FqCE}---$\mathcal{F}_{N\rightarrow\infty}^{\textrm{CE}}\!=\!\mathcal{F}_{\textrm{as}}^{\textrm{CE}}$.

What is more, when considering channels for which the asymptotic CE
method fails, as it is not possible to set $\beta_{\tilde{K}}\!=\!0$
in \eref{eq:FujiBound} for any Kraus representation, \eref{eq:FqCEN}
still applies; it is the finite-$N$ CE method that provides the correct
CE-based bound in the $N\!\rightarrow\!\infty$ limit that in principle
may then surpass the SQL-like scaling. On the other hand, when dealing
with estimation schemes in which one can moderate the effective amount
of loss (i.e. the form of $\Lambda_{\varphi}$) depending on the number
of particles, the asymptotic bound $\mathcal{F}_{\textrm{as}}^{\textrm{CE}}$
may not actually be the tightest within the CE method. The $\beta_{\tilde{K}}$
of \eref{eq:FujiBound} and \eref{eq:FqCEN} becomes then a function
of $N$ and it may not be asymptotically optimal to set it equal to
zero by imposing condition \eref{eq:FIC}. Yet the finite-$N$ CE
method, being not constrained with $\beta_{\tilde{K}}\!=\!0$, still
yields the correct CE-based bound on precision as $N\!\rightarrow\!\infty$.
This fact has been utilized in \cite{Chaves2012}, where, owing to
the $N$-dependence of $\beta_{\tilde{K}}$, the finite-$N$ CE method
provided an asymptotic bound indeed tighter than the naively calculated
$\mathcal{F}_{\textrm{as}}^{\textrm{CE}}$. What is more, the $\mathcal{F}_{N\rightarrow\infty}^{\textrm{CE}}$
has been numerically shown there to be saturable, what proves the
power of the more agile finite-$N$ CE method.
For phase estimation with various decoherence models including dephasing, depolarization, loss and spontaneous emission described
in detail in \ref{sec:ChannelsSpecs},
we observe that the  finite-$N$ CE bound is simply related to its asymptotic form as
\begin{equation}
\mathcal{F}_{N}^{\textrm{CE}}=\frac{N\,\mathcal{F}_{\textrm{as}}^{\textrm{CE}}}{N+\mathcal{F}_{\textrm{as}}^{\textrm{CE}}},
%\qquad
%\therefore\qquad N=1:\quad\mathcal{F}\!\left[\Lambda_\varphi\otimes\mathcal{I}\right]\!=
%\!\frac{\mathcal{F}_{\textrm{as}}^{\textrm{CE}}}{1+\mathcal{F}_{\textrm{as}}^{\textrm{CE}}},
\label{eq:F_CE_anal}
\end{equation}
where one should substitute for $\mathcal{F}_{\textrm{as}}^{\textrm{CE}}$
the corresponding asymptotic CE bounds presented in \Tref{tab:FqCh}
\footnote{In the case of spontanous emission noise the formula is valid only for $N\geq2$, what we suspect to be a consequence
of the spontaneous emission channel being an extremal map \cite{Bengtsson2006}.}.
%As a result, its Kraus representations do not allow for much freedom
%in the minimization over $\mathbf{h}$ in \eref{eq:FqCEN}. Hence,
%already for $N\!=\!2$ at which the second term in \eref{eq:FqCEN}
%is for the first time non-zero, the finite-$N$ CE bound may not be
%very tight.
\begin{figure}[!t]
\includegraphics[width=1\columnwidth]{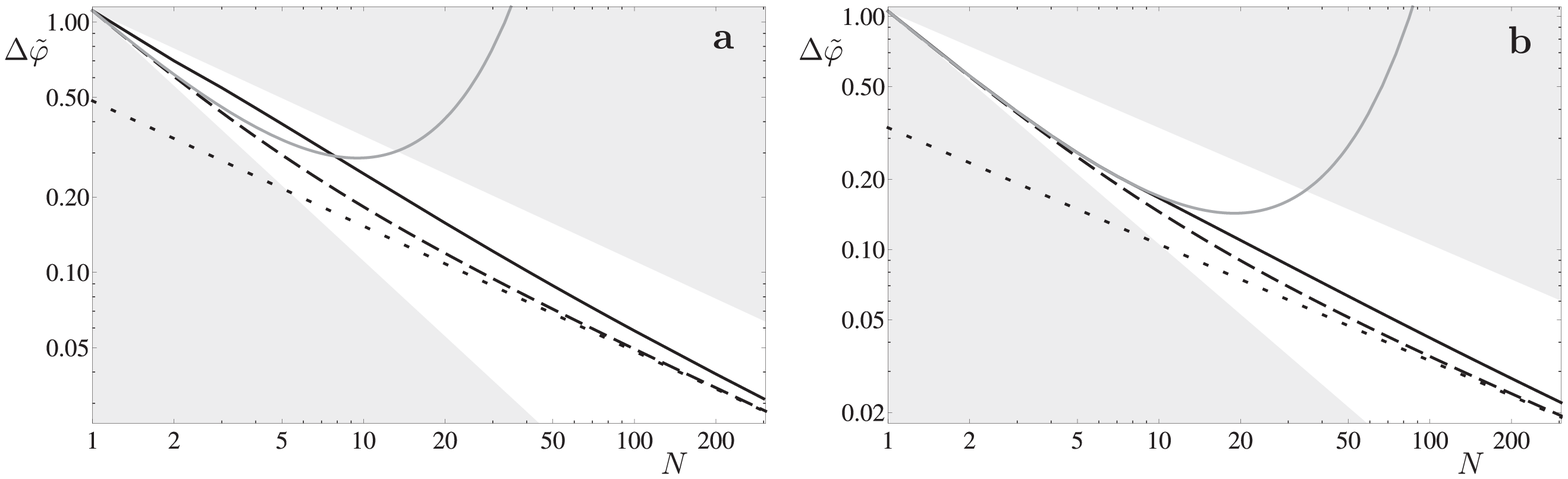}
\caption{\label{fig:plots}\textbf{Phase estimation CE method-based bounds on precision}
\protect \\
(\textbf{a}) \emph{Dephasing}: Finite-$N$ (\emph{dashed})
and the asymptotic CE bounds (\emph{dotted}) on estimation uncertainty
as compared with the precision achieved by utilizing spin-squeezed states in a Ramsey spectroscopy setup
(\emph{solid black}) and GHZ states with optimal measurement (\emph{solid grey}) for a probe consisting of $N$
atoms experiencing uncorrelated dephasing with $\eta=0.9$.
\protect \\
(\textbf{b}) \emph{Loss}: Lossy interferometry with particle survival probability $\eta=0.9$, e.g. Mach-Zehnder interferometer
experiencing photonic loss in both of its arms, with effective power transmission $\eta$. The smallest uncertainty in a phase estimation
scheme is quantified by calculating the QFI for numerically optimized
$N$-particle input states (\emph{solid black}) which only at low $N$ can be approximated by N00N states (\emph{solid grey}). Again, finite-$N$ (\emph{dashed}) and asymptotic
CE bounds (\emph{dotted}) on precision are shown for comparison.}
\end{figure}
For dephasing and loss decoherence models, we show explicitly
in \Fref{fig:plots} both the asymptotic and the finite-$N$ bounds
accompanied by the plots of actual precision achievable with explicit estimation strategies optimal either
in the small or large $N$ regime.

In the first case, depicted in \Fref{fig:plots}\textbf{a}, we consider a Ramsey spectroscopy
setup of \cite{Wineland1992,Bollinger1996} in which the probe consists
of atoms prepared in a spin-squeezed state \cite{Orgikh2001}.
The parameter is then encoded in the phase of a unitary rotation generated
by the total angular momentum of the atoms that simultaneously experience
uncorrelated dephasing. After measuring probe's total angular momentum
perpendicular to the one generating the estimated phase change, the
parameter is reconstructed with uncertainty plotted in \Fref{fig:plots}\textbf{a}.
For comparison, the maximal precision theoretically achievable with
Greenberger-Horne-Zeilinger (GHZ) \cite{GHZ} input states is also shown. The QFI for a GHZ-based strategy
is $\mathcal{F}_{N}^{\textrm{GHZ}}\!=\!\eta^{2N}N^2$ which for low $N$ attains
the finite-$N$ CE bound. This fact proves that in experiments with only few particles involved, such as
\cite{Leibfried2004}, it is optimal to use the GHZ states as inputs despite the uncorrelated
dephasing present.

In the second lossy interferometry case, shown in \Fref{fig:plots}\textbf{b}, each particle is subject to an independent loss process with
survival probability $\eta$, as in e.g. Mach-Zehnder interferometer
with effective power transmittance $\eta$ in both arms \cite{Dorner2009,Demkowicz2009}
that represents preparation, transmission and detection loss \cite{Kolodynski2010}.
Here the solid black line represents the maximal QFI achieved with
numerically optimised $N$-particle or equivalently $N$-photon states
\footnote{As it is optimal to consider indistinguishable, bosonic particles \cite{Demkowicz2009}.}.
As expected, it coincides for low $N$ with the QFI attained by
the so called N00N states \cite{Lee2002}, $\mathcal{F}_{N}^{\textrm{N00N}}\!=\!\eta^{N}N^2$,
which are the optical equivalents of the GHZ states previously considered. The plot indicates that
the finite-$N$ CE bound may be considered in this case to be tight only for moderate and very large $N$.
Although in the case of lossy optical interferometry, the maximal quantum enhancement of \Tref{tab:qEnh} can also be achieved
via an estimation strategy that employs squeezed-light as the input
with mean number of photons constrained to $N\!=\!\bar{N}$ \cite{Caves1981, Demkowicz2013},
we cannot compare its precision with the one bounded through $\mathcal{F}_{N=\bar{N}}^{\textrm{CE}}$.
As $N\!\cdot\!\mathcal{F}_{N}^{\textrm{CE}}$ is a convex quantity
in $N$, one cannot use it naively to constrain precision after replacing
$N$ by the mean number of photons $\bar{N}$. This contrasts the
case of the (constant) asymptotic CE bound, which yields $N\!\cdot\!\mathcal{F}_{\textrm{as}}^{\textrm{CE}}$
being linear in $N$, so that it also applies to estimation
strategies employing states of indefinite photon number, as pointed in \cite{Demkowicz2012,Demkowicz2013}.

%%%%%%%%%%%%%%%%%%%%%%%%%%%%%%%%%%%%%%%%%%%%%%%%%%%%%%%%%%%%%%%%%%%%%%%%%%%%%%%%%%%%%%%%%%%5

\section{Frequency estimation in atomic models \label{sec:FreqEstim}}

We apply the methods discussed above to the case of frequency estimation
problems in atomic spectroscopy. The general Ramsey spectroscopy setup
considered in \cite{Wineland1992,Bollinger1996,Andre2004,Auzinsh2004,Huelga1997,Shaji2007,Dorner2012,Chaves2012}
corresponds to $N$ identical two-level atoms---spin-$1/2$ systems---with 
their states separated, where typically the detuning $\omega$ of an external oscillator frequency
from the atoms transition frequency is to be estimated. We assume that the full experiment takes
an overall time $T$, during which the estimation procedure is repeated
$k\!=\! T/t$ times, where $t$ is the evolution duration of each
experimental shot. The quantum Cram\'{e}r-Rao bound \eref{eq:qCRB}
on precision of the estimate $\tilde{\omega}$ can be then conveniently
rewritten as
\begin{equation}
\Delta\tilde{\omega}\,\ge\frac{1}{\sqrt{T\, f_t \!\left[\rho_{\omega}^{N}(t)\right]}}\;,\label{eq:freqQCRB}
\end{equation}
where $f_t \!\left[\varrho(t)\right]\!=\! F_{\textrm{Q}}\!\left[\varrho(t)\right]/t$
is now the effective QFI per shot duration and $\rho_{\omega}^{N}(t)$
denotes the final state of the whole probe containing $N$ particles being measured in each shot. The total time
$T$ plays then the role of $k$ in \eref{eq:qCRB} and, after fixing
$t$, the bound \eref{eq:freqQCRB} can always be saturated as $T\!\rightarrow\!\infty$.
The evolution of the probe can be modelled by the master equation
of the Lindblad \cite{Breuer2002} form
\begin{equation}
\frac{\partial\rho_{\omega}^{N}(t)}{\partial t}=\sum_{n=1}^{N}\;\textrm{i}\,\frac{\omega}{2}\left[\sigma_{3}^{(n)},\rho_{\omega}^{N}(t)\right]+\mathcal{L}^{(n)}\!\left[\rho_{\omega}^{N}(t)\right],\label{eq:freqMasterEq}
\end{equation}
where $\sigma_{3}^{(n)}$ is the Pauli operator generating a unitary
rotation of the $n$'th atom around the $z$ axis in its Bloch ball
representation. The uncorrelated noise is represented by the Liouvillian
part $\mathcal{L}^{(n)}$ acting independently on each particle, here the $n$'th,
 so that effectively $\rho_{\omega}^{N}(t)\!=\!\Lambda_{\omega;t}^{\otimes N}\!\left[\left|\psi_{\textrm{in}}^N\right\rangle\right]$ with channel $\Lambda_{\omega;t}$
representing the overall single particle evolution over time $t$.
\begin{table}[!t]
\begin{tabular}{|>{\centering}m{2.35cm}|>{\centering}m{0.8cm}|>{\centering}m{1.5cm}|>{\centering}m{4.8cm}|>{\centering}m{0.55cm}||>{\centering}m{0.95cm}|>{\centering}m{1.6cm}|>{\centering}m{1.6cm}}
\cline{1-7}
\textbf{Noise model} & $\mathfrak{f}\!\left[\Lambda_{\omega}\right]$ & $\mathfrak{f}\!\left[\Lambda_{\omega}\!\otimes\!\mathcal{I}\right]$ & $\mathfrak{f}_{N}^{\textrm{CE}}$ {\small ($N\ge2$)} & $\mathfrak{f}_{\textrm{as}}^{\textrm{CE}}$ & $\chi\!\left[\Lambda_{\omega}\right]$ & $\!\chi\!\left[\Lambda_{\omega}\!\otimes\mathcal{I}\right]$ & \tabularnewline
\cline{1-7}
\emph{Dephasing} & $\frac{1}{2\,\textrm{e}\gamma}$ & $\frac{1}{2\,\textrm{e}\gamma}$ & $\frac{N}{2\gamma}\frac{w_{1}\!\left[N\right]}{1+\left(\textrm{e}^{w_{1}\!\left[N\right]}-1\right)N}$ & $\frac{1}{2\gamma}$ & {\scriptsize $=$}$\sqrt{\textrm{e}}$ & {\scriptsize $=$}$\sqrt{\textrm{e}}$ & ~\\
\tabularnewline
\cline{1-7}
\emph{\negthinspace{}\negthinspace{}Depolarization} & $\frac{3}{4\,\textrm{e}\gamma}$ & {\scriptsize $\approx\!1.27$}$\frac{3}{4\,\textrm{e}\gamma}$ & $\!\frac{3N}{4\gamma}\frac{\alpha\, w_{\beta}\left[N\right]}{2+\left(\textrm{e}^{\frac{\alpha}{4}w_{\beta}\left[N\right]}-1\right)\left(\textrm{e}^{\frac{\alpha}{4}w_{\beta}\left[N\right]}+2\right)N}$ & $\frac{1}{\gamma}$ & {\scriptsize $\le$}$\sqrt{\frac{4\textrm{e}}{3}}$ & {\scriptsize $\approx\!0.89$}$\sqrt{\frac{4\textrm{e}}{3}}$ & ~\\
\tabularnewline
\cline{1-7}
\emph{Loss} & $\frac{1}{\textrm{e}\gamma}$ & $\frac{1}{\textrm{e}\gamma}$ & $\frac{N}{\gamma}\frac{w_{1}\!\left[N\right]}{1+\left(\textrm{e}^{w_{1}\!\left[N\right]}-1\right)N}$ & $\frac{1}{\gamma}$ & {\scriptsize $=$}$\sqrt{\textrm{e}}$ & {\scriptsize $=$}$\sqrt{\textrm{e}}$ & ~\\
\tabularnewline
\cline{1-7}
\emph{Spontaneous emission} & $\frac{1}{\textrm{e}\gamma}$ & $\!\frac{4\tilde{w}}{\gamma\left(1+\textrm{e}^{\tilde{w}/2}\right)^{2}}$ & $\frac{N}{\gamma}\frac{4\,w_{4}\!\left[N\right]}{4+\left(\textrm{e}^{w_{4}\!\left[N\right]}-1\right)N}$ & $\frac{4}{\gamma}$ & {\scriptsize $\le$}$\sqrt{2\textrm{e}}$ & {\scriptsize $=$}$\frac{1+\textrm{e}^{\tilde{w}/2}}{\sqrt{\tilde{w}}}$ & \tabularnewline
\cline{1-7}
\end{tabular}\caption{\label{tab:freqQEnh} \textbf{QFIs, CE bounds and quantum enhancements
in frequency  estimation.} In frequency estimation tasks, the precision is maximized by adjusting
the single experimental shot duration $t$. The $t$-optimised (extended)
channel QFIs as well as their finite-$N$ and asymptotic CE bounds
are presented, where $w_{x}\!\left[N\right]=1\!+\! W\!\left[\frac{x-N}{\textrm{e}N}\right]$,
$\tilde{w}=1\!+\!2\, W\!\left[\frac{1}{2\sqrt{\textrm{e}}}\right]$
and $W\!\left[x\right]$ is the Lambert $W$ function. As in the case
of depolarizing channel not all solutions possess an analytical form, only their numerical approximations are shown with $\alpha\!\approx\!2.20$ and
$\beta\!\approx\!1.32$. Right of the double-line the maximal quantum precision enhancements
are listed for the maps considered. In the case of dephasing noise the ultimate
$\sqrt{\textrm{e}}$ factor has already been reported in \cite{Huelga1997,Escher2011}. For unextended depolarization
and spontaneous emission maps, the derived enhancement factors may
possibly be not achievable.}
\end{table}
To model the decoherence we consider the dephasing, depolarization, loss
and spontaneous emission maps, which corresponding Liouvillians can
be found in \ref{sec:ChannelsSpecs}. As the estimated parameter corresponds
now to $\omega\!=\!\varphi/t$, where $\varphi$ is the phase of the
unitary rotation, the QFI via a parameter change just rescales, so
that $f_t\!\left[\varrho_{\omega}\right]\!=\! F_{\textrm{Q}}\!\left[\varrho_{\omega}\right]/t\!=\! F_{\textrm{Q}}\!\left[\varrho_{\varphi}\right]t$.
Moreover, due to the commutativity of the unitary and the considered
decoherence maps, we can without loss of generality utilize the results
presented for them in the previous sections. Defining the channel
QFI for frequency estimation tasks similarly to \eref{eq:FqCh} as
\begin{equation}
\mathfrak{f}\!\left[\Lambda_{\omega}\right]=\max_{0\le t\le T}\max_{\psi_{\textrm{in}}}\, f_{t}\!\left[\Lambda_{\omega;t}\!\left[\left|\psi_{\textrm{in}}\right\rangle \right]\right]
\label{eq:FtCh}
\end{equation}
we can compute all the appropriate expressions for QFIs and the asymptotic
bounds of \Tref{tab:FqCh} as well as the finite-$N$ bounds of \eref{eq:F_CE_anal}
by substituting for the effective time dependence of the decoherence
strength $\eta(t)$, which is determined by the master equation \eref{eq:freqMasterEq}
(see \ref{sec:ChannelsSpecs}). Then, any quantity $\mathcal{F}$
listed in \Tref{tab:FqCh} transforms to its $t$-optimized equivalent
as $\mathfrak{f}\!=\!\underset{0\le t\le T}{\max}\,\mathcal{F}\, t$.
In \Tref{tab:freqQEnh} we present the channel QFIs relevant for frequency
estimation, their asymptotic and finite-$N$ CE bounds, as well as
the maximal quantum precision enhancements for each model considered.
In the case of dephasing, we recover the results of \cite{Huelga1997,Escher2011,Chaves2012},
whereas for depolarizing, loss and spontaneous emission maps we obtain the
QFIs and their bounds, which to our knowledge have not been reported
in the literature before. However, similarly to the case of quantum phase estimation
summarized in \Tref{tab:qEnh}, the obtained quantum enhancement factors for
depolarization and spontaneous emission channels serve only as bounds, as they are not guaranteed to be saturable.

%%%%%%%%%%%%%%%%%%%%%%%%%%%%%%%%%%%%%%%%%%%%%%%%%%%%%%%%%%%%%%%%%%%%%%%%%%%%%%%%%%%%%%%%%%%

\section{Estimation of decoherence strength \label{sec:LossEstim}}

Lastly, we would like to emphasize that the CS, QS and CE methods
described in \Sref{sec:EstNChannels} also apply to estimation tasks
in which the estimated parameter is not encoded in the unitary, noiseless
part of the system evolution. Examples of such schemes are the experimentally
motivated ones, in which one tries to quantify the effective strength
of noise or loss present in the apparatus. That is why we consider
again the channels described in \ref{sec:ChannelsSpecs}, but this
time with the parameter to be estimated being the decoherence strength
$\eta$. This kind of problems has been widely considered not only
in the estimation theory \cite{Monras2007,Adesso2009,Crowley2012,Hotta2005},
but also when examining issues of channel discrimination \cite{Sacchi2005,DAriano2005a}
with particular application in quantum reading \cite{Pirandola2011,Pirandola2011a,Nair2011a}.
As compared to unitary rotations, the nature of the estimated parameter
is dramatically different. In unitary parameter case, the use of entangled input state of $N$
particles results in an effective $N$-times higher ``angular speed'' of rotation leading
to the HL in the absence of noise. In decoherence strength estimation
tasks, a change in the parameter value can be geometrically interpreted
in the space of all valid quantum channels as a ``movement''
in the direction away from the boundary of the space of the relevant
CPTP maps, for which ``speed''  cannot be naively amplified $N$-times when employing
$N$ parallel channels. Hence, as in the case of lossy unitary
rotation estimation, the optimal entangled inputs must lead not to scaling but constant factor quantum enhancements,
which again can be quantified by the methods of \Sref{sec:EstNChannels}. This also explains
that for all the four noise models considered, the purely geometrical notion of
classical simulability is enough to bound most tightly the maximal asymptotic precision of
estimation. However, as for them also $\mathcal{F}_{\textrm{as}}^{\textrm{CS}}\!=\!\mathcal{F}_{\textrm{as}}^{\textrm{QS}}\!=\!\mathcal{F}_{\textrm{as}}^{\textrm{CE}}\!=\!\mathcal{F}\left[\Lambda_{\eta}\otimes\mathcal{I}\right]$,
the CS-based asymptotic quantum enhancement corresponds to the classical estimation strategy that employs
independent but extended channels.
The fact that factorizable inputs---uncorrelated in between the extended $N$ channels but possibly requiring
entanglement between each single particle and its ancilla---are optimal for noise estimation
with extended channels, has already been noticed for the low-noise \cite{Hotta2006} and
generalized Pauli \cite{Fujiwara2003} channels, of which the latter contain the
dephasing and depolarization maps studied here.

\begin{table}[!t]
\begin{tabular}{|>{\centering}m{4.5cm}|>{\centering}m{2.5cm}|>{\centering}m{3cm}||>{\centering}m{3.5cm}|>{\centering}m{1.6cm}}
\cline{1-4}
\textbf{Channel considered} & $\!\mathcal{F}\!\left[\Lambda_{\eta}\right]$ & $\!\mathcal{F}\!\left[\Lambda_{\eta}\!\otimes\!\mathcal{I}\right]$ & \emph{\negthinspace{}}$\mathcal{F}_{\textrm{as}}^{\textrm{CS}}\!=\!\mathcal{F}_{\textrm{as}}^{\textrm{QS}}\!=\!\mathcal{F}_{\textrm{as}}^{\textrm{CE}}$  & ~\\
\tabularnewline
\cline{1-4}
\emph{Dephasing} & $\frac{1}{1-\eta^{2}}$ & $\frac{1}{1-\eta^{2}}$ & $\;\frac{1}{1-\eta^{2}}${\footnotesize{} \cite{Fujiwara2003}} & ~\\
\tabularnewline
\cline{1-4}
\emph{\negthinspace{}Depolarization} & $\;\;\frac{1}{1-\eta^{2}}$ {\footnotesize \cite{Fujiwara2001}} & $\;\;\frac{3}{\left(1-\eta\right)\left(1+3\eta\right)}$ {\footnotesize \cite{Fujiwara2001}} & $\;\;\frac{3}{\left(1-\eta\right)\left(1+3\eta\right)}$ {\footnotesize \cite{Fujiwara2008,Fujiwara2003}} & ~\\
\tabularnewline
\cline{1-4}
\emph{Loss} & $\frac{1}{\eta\left(1-\eta\right)}$ & $\frac{1}{\eta\left(1-\eta\right)}$ & $\frac{1}{\eta\left(1-\eta\right)}$ & ~\\
\tabularnewline
\cline{1-4}
\emph{Spontaneous emission} & $\;\;\frac{1}{\eta\left(1-\eta\right)}$ {\footnotesize \cite{Fujiwara2004}} & $\;\;\frac{1}{\eta\left(1-\eta\right)}$ {\footnotesize \cite{Fujiwara2004}} & $\frac{1}{\eta\left(1-\eta\right)}$ & ~\\
\tabularnewline
\cline{1-4}
\end{tabular}\caption{\label{tab:FqCh_loss} \textbf{Decoherence strength estimation quantified via channel QFIs and their asymptotic bounds.}
Definitions of channels listed in the first column can
be found in \ref{sec:ChannelsSpecs}. In contrast
to phase estimation examples given in \Tref{tab:FqCh}, the variable to be estimated here is the decoherence parameter $\eta$,
 $(0\!\le\!\eta\!<\!1)$. Due
to the different nature of the estimated parameter, the geometrical
CS method provides bounds that not just most
tightly limit the asymptotic extended channel QFIs, but actually coincide
with its value. The results prove that only in the case of depolarization
channel the precision can be enhanced with the use of quantum estimation
strategies, as for all other cases
$\mathcal{F}\!\left[\Lambda_{\eta}\right]\!=\!\mathcal{F}\!\left[\Lambda_{\eta}\otimes\mathcal{I}\right]=\mathcal{F}_{\textrm{as}}$. }
\end{table}

In the case of dephasing channel, we further realize that the extension at the single channel level
is also unnecessary, as $\mathcal{F}\left[\Lambda_{\eta}\right]\!=\!\mathcal{F}\left[\Lambda_{\eta}\otimes\mathcal{I}\right]\!=\!1/(1-\eta^{2})$,
and the geometrically dictated bound of CS is attainable classically
just by employing unentangled qubits in any pure state lying on the Bloch
sphere equatorial plane. Similarly, the spontaneous emission and loss maps also turn
out to be fully classical. In the first case, the asymptotic CS bound coincides with the extended and unextended channel QFIs
derived in \cite{Fujiwara2004}, whereas for the loss channel we obtain
$\mathcal{F}_{\textrm{as}}^{\textrm{CS}}\!=\!\mathcal{F}\left[\Lambda_{\eta}\right]\!=\!\mathcal{F}\left[\Lambda_{\eta}\otimes\mathcal{I}\right]\!=\!1/(\eta(1-\eta))$,
which at the single channel level is achieved by a photon in any mixed state with the ancilla being redundant. On the one hand, this emphasizes that entanglement
between the photons entering the interferometer is unnecessary and agrees with the results of  \cite{Monras2007,Adesso2009}
confirming that the total photon number fluctuations are really the ones that limit the precision.
These can be reduced by employing Gaussian states \cite{Monras2007} or in principle
fully eliminated by the use of Fock states \cite{Adesso2009} that attain the CS-bound.
%\comm{[TO DO:]} On the other hand, the fact that the channel extension at the single photon level does not improve the precision agrees with the
%channel discrimation tasks \cite{Pirandola2011,Pirandola2011a,Nair2011a}, for which it has been shown that additional ancillae (the idler mode)
%are not necessary when distinguishing channels infinitesimally close to each other
%in the space all quantum maps
%what is equivalent to the local estimation considered in this paper. Yet, in \cite{Nair2011} it has been proven that the idler mode makes it possible %to measure at
%the interferometer output ports how many photons were lost in each of its arms, what in principle may improve the precision \cite{Demkowicz2009}. %\comm{This, however, also agrees with the single photon channel notion, as ....
%distinguishability??}

The case of depolarization map is different, as it is known that for qubits \cite{Fujiwara2001,Frey2011} the precision of estimation may
be improved by extending the channel, i.e. $\mathcal{F}\left[\Lambda_{\eta}\right]\!<\!\mathcal{F}\left[\Lambda_{\eta}\otimes\mathcal{I}\right]$.
This leaves the space for possible enhancement thanks to the use of entangled probes \emph{between} \emph{unextended} channels
and indeed this fact has been observed already when considering two depolarization channels used in parallel
\cite{Fujiwara2001}. The results are summarized in Table \ref{tab:FqCh_loss}.
%in which for $\eta\!>\!1/\sqrt{3}$ by the use of maximally entangled inputs the precision is optimally improved.
%Hence, the CS-bound $\mathcal{F}_{\textrm{as}}^{\textrm{CS}}\!=\!\mathcal{F}\left[\Lambda_{\eta}\otimes\mathcal{I}\right]$
%applied to strategies with no ancillae involved, if tight, may be attained only by employing
%some entangled input states of the probe. We seek for the candidates by firstly utilizing the work of \cite{Simon2002}, and generalizing the result
%of \cite{Fujiwara2001} to GHZ input states of arbitrary number of particles, $\left|\textrm{GHZ}\right\rangle _{N}$, that undergo uncorrelated %depolarization.
%For those we obtain the precision scaling shown in \Fref{fig:dep_est} which reaches its maximum at $N\!=\!2$ going back to classical scaling, which is
%attained asymptotically. Hence, if only GHZ states were allowed, as $F_{\textrm{Q}}\!\left[\Lambda_{\eta}\!\left[\left|\textrm{GHZ}\right\rangle _{2}\right]\right]/2\!>\! F_{\textrm{Q}}\!\left[\Lambda_{\eta}\!\left[\left|\textrm{GHZ}\right\rangle _{3}\right]\right]/3$,
%it is always optimal to group channels into $N/2$ pairs, to which any maximally entangled states, e.g. $\left|\textrm{GHZ}\right\rangle _{2}$, should be inputted.
%\comm{[have numerics going towards the bound. find a good candidate...]}

%%%%%%%%%%%%%%%%%%%%%%%%%%%%%%%%%%%%%%%%%%%%%%%%%%%%%%%%%%%%%%%%%%%%%%%%%%%%%%%%%%%%%%%%%%%5

\section{Further discussion \label{sec:FurDisc}}

We would also like to point that the SQL-like bounds, universally valid
in practical metrological scenarios, allow one to avoid some of the
controversies characteristic for idealized decoherence-free scenarios.
When decoherence is \emph{not} present and the probe states with indefinite number of particles are considered, such as e.g. squeezed states of light,
the exact form of HL needs to be reconsidered \cite{Hyllus2010, Hofmann2009, Zwierz2010} since the
direct replacement of $N$ with mean number of particles $\bar{N}$ may make the HL invalid.
Moreover, the final claims on the achievable precision scaling may strongly depend on the form
of a priori parameter knowledge assumed, and lead to some apparent contradictions \cite{Anisimov2010, Giovannetti2012}.
These difficulties do not arise in the realistic metrological schemes, as
the asymptotic SQL-like bounds are valid also when $N$ is replaced by $\bar{N}$ for indefinite particle number state
\cite{Demkowicz2012, Demkowicz2013}. The bounds
derived in the \emph{local approach} (small parameter fluctuation)
 based on the calculation of the QFI are saturable
 in a single-shot scenario unlike the decoherence-free case when only
 after some number of independently repeated experiments one may expect to
 approach the theoretical limits \cite{Pezze2008,Giovannetti2012a}.
This is due to the fact that by employing input states of
grouped particles, which possess no correlations in between the groupings,
and by letting the groups to be of finite but sufficiently large size,
one can attain the ultimate asymptotic SQL bound up to any precision.
Since saturability of the QFI bounds for independently prepared probes
is well established \cite{Helstrom1976, Holevo1982, Braunstein1994, Kahn2009},
the operational meaning of the QFI is also clear in the single shot scenario.
The above argument also suggests the asymptotically optimal form of the input states, which should
include ones that do not possess long-range correlations in between the particles.
This observation has already been made in \cite{Sorensen2001} and indicates that in
methods designed to search for the optimal inputs in scenarios with uncorrelated noise
one may restrict himself to states with short-range correlations such as for example
the matrix product states of low bond dimensions \cite{Jarzyna2013}. 

We also conjecture from the point of view of the asymptotic SQL-like bounds
that the specific form of the \emph{a priori} knowledge should not play an important role.
In particular, we expect that various methods such as \emph{Bayesian} \cite{Chiribella2005,Boixo2008,Teklu2009,Demkowicz2011}
or \emph{information theoretic} \cite{Hall2012,Nair2012} should recover
the bounds compatible with the ones obtained via the \emph{local approach} considered in this paper.
This statement is known to hold in the case of optical interferometry with losses \cite{Kolodynski2010,Knysh2011}; it is an intriguing
question whether analogous claims can be made in more general scenarios.

\section{Summary and outlook \label{sec:SumOut}}
We have constructed explicit methods capable of determining fundamental bounds 
on quantum enhancement in metrological setups in presence of uncorrelated noise. The methods are based on the
study of the structure of a single-particle quantum channel that represents the decoherence process. The methods 
do not require any kind of educated guess, nor an involved numerical optimization---given a set of Kraus operators
representing the channel, bounds on precision can be derived immediately without the need to search e.g. for the optimal input states. 
   We have discussed the efficiency of CS, QS, RLD and CE methods in providing asymptotic bounds on precision for phase estimation under a number of different decoherence models.
We have also generalized the CE method in order to provide tighter bounds in the regime of finite number of particles and we have showed that this generalization can be again cast in the form of a semi-definte program. The methods have also been applied to a related problem of
frequency estimation. Moreover, it has been shown that when thinking of estimation of the decoherence parameter itself 
already the simplest approach based on the CS method typically provides the tightest bounds. 
While the methods are efficient as they avoid the search for the optimal many particle input states, 
formulation of an explicit optimal estimation strategy may in general require performing such a search. 
Hopefully, the optimal states are expected to have a relatively simple structure and 
can be searched within a restricted class of states such as e.g. squeezed or matrix  product states \cite{Sorensen2001, Orgikh2001, Jarzyna2013, Demkowicz2013}. Once the precision calculated for a given input state hits the fundamental bound one is guaranteed that 
the optimal strategy has been identified. The natural future work on our methods is to generalize them and study their applicability in the multi-parameter estimation schemes where it is \emph{a priori} not clear which of the different approaches will be the most fruitful and whether
some nontrivial new bounds may be derived.

%%%%%%%%%%%%%%%%%%%%%%%%%%%%%%%%%%%%%%%%%%%%%%%%%%%%%%%%%%%%%%%%%%%%%%%%%%%%%%%%%%%%%%%%%%%5
\ack{}{}
\addcontentsline{toc}{section}{Acknowledgements}

We would like to thank M\u{a}d\u{a}lin Gu\c{t}\u{a} and Lorenzo Maccone for valuable feedback
as well as Konrad Banaszek for constant support. J.K. also acknowledges Michal Sedl\'{a}k and Jonatan
Bohr Brask for helpful comments and thanks Marcin Jarzyna for many
fruitful discussions as well as the data utilized
in \Fref{fig:plots}. This research was supported by Polish NCBiR under the ERA-NET CHIST-ERA project QUASAR,
Foundation for Polish Science TEAM project co-financed by the EU European Regional
Development Fund and FP7 IP project Q-ESSENCE.

%%%%%%%%%%%%%%%%%%%%%%%%%%%%%%%%%%%%%%%%%%%%%%%%%%%%%%%%%%%%%%%%%%%%%%%%%%%%%%%%%%%%%%%%%%%5
\appendix
%\addcontentsline{toc}{section}{Appendices}
%%
\addtocontents{toc}{\setlength{\cftsecnumwidth}{16ex}}
\addtocontents{toc}{\setlength{\cftsubsecnumwidth}{16ex}}

\section{\label{sec:ChannelsSpecs}Channels considered}

We adopt the standard notation in which $\mathbb{I}_{d}$ represents
a $d\times d$ identity matrix and $\left\{ \sigma_{i}\right\} _{i=1}^{3}$
are the Pauli operators. In \Sref{sub:UnitWithDeco} parameter $\varphi$
to be estimated is the rotation angle around the $z$ axis of the
Bloch ball generated by the unitary operator $U_{\varphi}\!=\!\exp\!\left[\textrm{i}\sigma_{3}\varphi/2\right]$.
We consider maps $\mathcal{D}_{\eta}$ with decoherence parameter $\eta$
that commute with such rotation, whence $\Lambda_{\varphi}\!\left[\varrho\right]\!=\!\mathcal{D}_{\eta}\!\left[U_{\varphi}\varrho U_{\varphi}^{\dagger}\right]\!=\! U_{\varphi}\mathcal{D}_{\eta}\!\left[\varrho\right]U_{\varphi}^{\dagger}$,
and are defined accordingly by the Kraus operators presented below.
For each case, we also specify the purification determining the extended
channel QFI \eref{eq:FqChExt} ($\mathcal{F}\!\left[\Lambda_{\varphi}\otimes\mathcal{I}\right]$
in \Tref{tab:FqCh}) in the form of the optimal generator of Kraus
representation rotation $h$, as introduced in \eref{eq:FqChPurif}.
Dealing with frequency estimation tasks discussed in \Sref{sec:FreqEstim}
we construct the effective one-particle Kraus operators by substituting
$\varphi\!\rightarrow\!\omega t$ and $\eta\!\rightarrow\!\eta(t)$
in the nominal ones, where $\omega$ is the estimated frequency detuning.
For all models, we explicitly
write the Liouvillian $\mathcal{L}^{(n)}$ determining the noise affecting
each particle, see \eref{eq:freqMasterEq}, and the effective form
of $\eta(t)$. When discussing decoherence strength estimation in
\Sref{sec:LossEstim}, we consider solely each of the following noise
maps with $\eta$ being now the parameter to be estimated: $\Lambda_{\varphi=\eta}\!=\!\mathcal{D}_{\eta}$.

\subsection{Dephasing}
\begin{itemize}
\item \emph{Decoherence parameter}, $\eta$, represents the final
equatorial radius of the Bloch ball shrank uniformly in the $xy$
plane by the channel.
\item \emph{Kraus operators}:
\begin{equation}
K_{1}=\sqrt{\frac{1+\eta}{2}}\;\mathbb{I}_{2}\,,\quad K_{2}=\sqrt{\frac{1-\eta}{2}}\;\sigma_{3}\,.
\end{equation}

\item \emph{Optimal purification }determining\emph{ }the extended channel
QFI\emph{ }\eref{eq:FqChExt}:
\begin{equation}
h=\frac{\sqrt{1-\eta^{2}}}{2}\;\sigma_{1}\,.
\end{equation}

\item \emph{One-particle Liouvillian }for frequency estimation tasks:
\begin{equation}
\mathcal{L}^{(n)}\!\left[\varrho\right]=\frac{\gamma}{2}\left(\sigma_{3}^{(n)}\varrho\,\sigma_{3}^{(n)}-\varrho\right)\qquad\therefore\quad\eta(t)=\textrm{e}^{-\gamma t}.
\end{equation}

\end{itemize}

\subsection{Depolarization}
\begin{itemize}
\item \emph{Decoherence parameter}, $\eta$, represents the final
radius of the Bloch ball shrunk isotropically by the channel.
\item \emph{Kraus operators}:
\begin{equation}
K_{1}=\sqrt{\frac{1+3\eta}{4}}\;\mathbb{I}_{2}\,,\quad\left\{ K_{i}=\sqrt{\frac{1-\eta}{4}}\,\sigma_{i-1}\right\} _{i=2\dots4}\!\!\!\!\!\!\!\!\!\!\!\!\!.
\end{equation}

\item \emph{Optimal purification }determining the extended channel QFI\emph{
}\eref{eq:FqChExt}:
\begin{equation}
h=\frac{1}{2}\left(\!\!\!\!\begin{array}{ccc}
0 & \!\!\!\!\!0\;\;0 & \!\!\!\!\!\xi\\
\begin{array}{c}
0\\
0
\end{array} & \!\!\!\!\!\left[\;\overset{\;}{\underset{\;}{\sigma_{2}}\;}\right] & \!\!\!\!\!\begin{array}{c}
0\\
0
\end{array}\\
\xi & \!\!\!\!\!0\;\;0 & \!\!\!\!\!0
\end{array}\!\!\!\!\right)\quad\textrm{with}\quad\xi=\frac{\sqrt{\left(1+3\eta\right)\left(1-\eta\right)}}{1+\eta}.
\end{equation}

\item \emph{One-particle Liouvillian }for frequency estimation tasks:
\begin{equation}
\mathcal{L}^{(n)}\!\left[\varrho\right]=\frac{\gamma}{2}\left(\frac{1}{3}\sum_{i=1}^{3}\sigma_{i}^{(n)}\varrho\,\sigma_{i}^{(n)}-\varrho\right)\qquad\therefore\quad\eta(t)=\textrm{e}^{-\frac{2\gamma}{3}t}.
\end{equation}

\end{itemize}

\subsection{Loss}
\begin{itemize}
\item \emph{Decoherence parameter}, $\eta$, represents survival probability of each of the particles that are subject
to independent loss processes. The channel on a single particle is formally a
map from a two- to a three-dimensional system with the output's third
dimension corresponding to the vacuum mode responsible for the particle
loss. Although in this case one should strictly write $\Lambda_{\varphi}\!=\!\mathcal{D}_{\eta}\!\left[U_{\varphi}\varrho U_{\varphi}^{\dagger}\right]\!=\!\tilde{U}_{\varphi}\,\mathcal{D}_{\eta}\!\left[\varrho\right]\tilde{U}_{\varphi}^{\dagger}$
with $\tilde{U}_{\varphi}$ acting on a different Hilbert space, the
effects of $U_{\varphi}$ and $\tilde{U}_{\varphi}$ are physically
indistinguishable, as the particle losses commute with the acquired
phase (for instance see \cite{Demkowicz2009}). In the case of optical interferometry,
$\eta$ represents the effective power transmittance assumed to be equal
in both arms and accounts for preparation and transmission loss as well as
detector inefficiencies in a Mach-Zehnder setup \cite{Kolodynski2010}.
\item \emph{Kraus operators}:
\begin{equation}
\!\!\!\!\!\!\!\!\!\!\!\!\!\!\!\!\!\!\!\!\!\!\!\! K_{1}=\left(\begin{array}{cc}
0 & 0\\
0 & 0\\
0 & \sqrt{1-\eta}
\end{array}\!\!\right)\!,\; K_{2}=\left(\!\!\begin{array}{cc}
0 & 0\\
0 & 0\\
\sqrt{1-\eta} & 0
\end{array}\right)\!,\; K_{3}=\left(\!\begin{array}{cc}
\sqrt{\eta} & 0\\
0 & \sqrt{\eta}\\
0 & 0
\end{array}\right)\!.
\end{equation}

\item \emph{Optimal purification }determining the extended channel QFI\emph{
}\eref{eq:FqChExt}:
\begin{equation}
h=-\frac{1}{2}\left(\!\!\!\begin{array}{cc}
\left[\;\overset{\;}{\underset{\;}{\sigma_{3}}\;}\right] & \!\!\!\begin{array}{c}
0\\
0
\end{array}\\
0\;\,\;0 & \!\!\!0
\end{array}\right)\!.
\end{equation}

\item \emph{One-particle Liouvillian }for the frequency estimation tasks:
\begin{equation}
\!\!\!\!\!\!\!\!\!\!\!\!\!\!\!\!\!\!\!\!\!\!\!\!\!\!\!\!\mathcal{L}^{(n)}\!\left[\varrho\right]=\gamma\sum_{m=0}^{1}\left(\sigma_{m,+}^{(n)}\varrho\,\sigma_{m,-}^{(n)}-\frac{1}{2}\left\{ \sigma_{m,-}^{(n)}\sigma_{m,+}^{(n)},\varrho\right\} \right)\qquad\therefore\quad\eta(t)=\textrm{e}^{-\gamma t},
\end{equation}
where $\sigma_{m,+}^{(n)}\!=\!\ket{\textrm{vac}}\!\bra{m}$ are the generators of transition to the vacuum mode from qubit basis states $\ket{0}$ and $\ket{1}$, such that $\sigma_{m,-}^{(n)}\!=\!\sigma_{m,+}^{(n)\dagger}$.

\end{itemize}
The methods discussed in the paper may also be easily applied to more general loss models such as: unequal loss in the two arms of an interferometer \cite{Dorner2009} or the models with distinguished preparation, transmission and detection loss \cite{Datta2011}. For the conciseness of the paper,
however, we restrict ourselves to the simplest loss model described above.

\subsection{Spontaneous emission (amplitude damping)}
\begin{itemize}
\item \emph{Decoherence parameter}, $\eta$, represents the radius
of the disk obtained by projecting the deformed Bloch ball outputted
by the channel onto the $xy$ plane.
\item \emph{Kraus operators}:
\begin{equation}
\!\!\!\!\!\!\!\!\!\!\!\!\! K_{1}=\left(\!\!\begin{array}{cc}
1 & 0\\
0 & \sqrt{\eta}
\end{array}\!\!\right),\quad K_{2}=\left(\!\!\begin{array}{cc}
0 & \sqrt{1-\eta}\\
0 & 0
\end{array}\!\!\!\right)\!.
\end{equation}

\item \emph{Optimal purification }determining\emph{ }the extended channel
QFI\emph{ }\eref{eq:FqChExt}:
\begin{equation}
\!\!\!\!\!\!\!\!\!\!\!\!\!h=\frac{1}{2}\left(\!\!\begin{array}{cc}
\xi & 0\\
0 & -1
\end{array}\!\!\right)\quad\textrm{with}\quad\xi=\frac{1-\sqrt{\eta}}{1+\sqrt{\eta}}\,.
\end{equation}

\item \emph{One-particle Liouvillian }for the frequency estimation tasks
($\sigma_{\pm}\!=\!\frac{1}{2}\left(\sigma_{1}\pm\textrm{i}\sigma_{2}\right)$):
\begin{equation}
\!\!\!\!\!\!\!\!\!\!\!\!\!\mathcal{L}^{(n)}\!\left[\varrho\right]=\gamma\left(\sigma_{+}^{(n)}\varrho\,\sigma_{-}^{(n)}-\frac{1}{2}\left\{ \sigma_{-}^{(n)}\sigma_{+}^{(n)},\varrho\right\} \right)\qquad\therefore\quad\eta(t)=\textrm{e}^{-\gamma t}.
\end{equation}

\end{itemize}
%%%%%%%%%%%%%%%%%%%%%%%%%%%%%%%%%%%%%%%%%%%%%%%%%%%%%%%%%%%%%%%%%%%%%%%%%%%%%%%%%%%%%%%%%%%5

\section{\label{sec:RLDEquivCS}Equivalence of RLD-based bound applicability and local classical simulability of a channel}

Given a channel---a CPTP map $\Lambda_{\varphi}\!:\mathcal{H}_{\textrm{in}}\!\rightarrow\!\mathcal{H}_{\textrm{out}}$---we 
define its C-J matrix representation \cite{Bengtsson2006}
as $\Omega_{\Lambda_{\varphi}}\!=\!\Lambda_{\varphi}\otimes\mathcal{I}\left[\left|\mathbb{I}\right\rangle \!\left\langle \mathbb{I}\right|\right]\!=\!\sum_{i}\left|K_{i}(\varphi)\right\rangle \!\left\langle K_{i}(\varphi)\right|$,
where $\left\{ K_{i}(\varphi)\right\} _{i=1}^{r}$ are $r$ linearly
independent Kraus operators of $\Lambda_{\varphi}$; we adopt a concise
notation for bipartite states, in which $\left|\phi\right\rangle \!=\!\sum_{i,j=1}^{\dim\mathcal{H}_{in}}\!\left\langle i\right|\!\phi\!\left|j\right\rangle \left|i\right\rangle \left|j\right\rangle \!=\!\phi\otimes\mathbb{I}\left|\mathbb{I}\right\rangle \!=\!\mathbb{I}\otimes\phi^{T}\left|\mathbb{I}\right\rangle $
with $\left|\mathbb{I}\right\rangle \!=\!\sum_{i=1}^{\dim\mathcal{H}_{in}}\!\left|i\right\rangle \left|i\right\rangle $.
For simplicity, from now onwards we drop the explicit $\varphi$ dependence
of operators, assuming that the estimation is performed for small
variations $\delta\varphi$ around a given, fixed $\varphi$. In Sup.
Mat. of \cite{Demkowicz2012} (Equation (S9)) it has been proven
that the condition for any channel to be $\varphi$-non-extremal
%(classically simulable)
at $\varphi$ is equivalent to the statement that there
exists a non-zero Hermitian matrix $\mu_{ij}$ such that
\begin{equation}
\dot{\Omega}_{\Lambda_{\varphi}}=\sum_{ij}\mu_{ij}\left|K_{i}\right\rangle \!\left\langle K_{j}\right|\!.\label{eq:CScond}
\end{equation}
On the other hand, the RLD-based bound exists there if and only if
\cite{Hayashi2011}
\begin{equation}
P_{\Omega_{\perp}}\dot{\Omega}_{\Lambda_{\varphi}}^{2}P_{\Omega_{\perp}}=0\label{eq:RLDcond}
\end{equation}
where $P_{\Omega_{\perp}}$ is the projection onto the null-space
$\Omega_{\perp}$, i.e. the subspace orthogonal to $\Omega_{\Lambda_{\varphi}}$,
so that $\forall_{i}\!:\, P_{\Omega_{\perp}}\left|K_{i}\right\rangle \!=\!0$.
The \eref{eq:CScond} implies \eref{eq:RLDcond}, as by substitution
\begin{equation}
\!\!\!\!\!\!\!\!\!\!\!\!\!\!\!\!\!\!\!\!\!\!\!\!\!\!\!\!\!\!\!\!\!\!\!\! P_{\Omega_{\perp}}\!\left(\!\sum_{ij}\mu_{ij}\left|K_{i}\right\rangle \!\left\langle K_{j}\right|\!\right)^{2}\!\! P_{\Omega_{\perp}}=\sum_{ij}\!\left(\!\sum_{p}\mu_{ip}\left\langle K_{p}|K_{p}\right\rangle \mu_{pj}\!\right)P_{\Omega_{\perp}}\left|K_{i}\right\rangle \!\left\langle K_{j}\right|P_{\Omega_{\perp}}=0\,,
\end{equation}
thus any $\varphi$-non-extremal channel must admit an RLD-based bound
on its extended QFI. In order to prove the other direction, we split
the derivatives of each C-J eigenvector into components supported
by $\Omega_{\Lambda_{\varphi}}$ and in the null-space $\Omega_{\perp}$:
$\left|\dot{K}_{i}\right\rangle \!=\!\sum_{j}\nu_{ij}\left|K_{j}\right\rangle \!+\!\left|K_{i}^{\perp}\right\rangle $.
Hence, after substituting for $\dot{\Omega}_{\Lambda_{\varphi}}$
the \eref{eq:RLDcond} then simplifies to
\begin{equation}
\left(\sum_{i}\left|K_{i}^{\perp}\right\rangle \!\left\langle K_{i}\right|\right)\left(\sum_{j}\left|K_{j}\right\rangle \!\left\langle K_{j}^{\perp}\right|\right)=0,
\end{equation}
and since $A^{\dagger}A\!=\!0$ implies $A^{\dagger}\!=\! A\!=\!0$
and $\left\{ \left|K_{i}\right\rangle \right\} _{i}$ are orthogonal,
we conclude that all $\left|K_{i}^{\perp}\right\rangle \!=\!0$. Thus,
\eref{eq:RLDcond} implies that $\left|\dot{K}_{i}\right\rangle \!=\!\sum_{j}\nu_{ij}\left|K_{j}\right\rangle $,
which due to the local ambiguity of Kraus representations \eref{eq:KrausReps}
is equivalent to $\left|\dot{\tilde{K}}_{i}\right\rangle \!=\!\sum_{j}\!\left(\nu_{ij}\!-\!\textrm{i}\,h_{ij}\right)\!\left|\tilde{K}_{j}\right\rangle $
for any Hermitian $h$. Therefore, without loss
of generality, we may set $h\!=\!-\frac{\textrm{i}}{2}\nu^{\textrm{AH}}$
after splitting $\nu$ into its Hermitian and anti-Hermitian parts
$\nu\!=\!\nu^{\textrm{H}}\!+\!\nu^{\textrm{AH}}$, so that $\left|\dot{\tilde{K}}_{i}\right\rangle \!=\!\sum_{j}\nu_{ij}^{\textrm{H}}\left|\tilde{K}_{j}\right\rangle $
with $\nu^{\textrm{H}}\!\ne\!0$ for any non-trivial channel. Finally,
we can write
\begin{equation}
\dot{\Omega}_{\varphi}=\sum_{i}\left|\dot{\tilde{K}}_{i}\right\rangle \!\left\langle \tilde{K}_{i}\right|+\left|\tilde{K}_{i}\right\rangle \!\left\langle \dot{\tilde{K}}_{i}\right|=2\sum_{ij}\nu_{ji}^{\textrm{H}}\left|\tilde{K}_{i}\right\rangle \!\left\langle \tilde{K}_{j}\right|\label{eq:RLDcondFin}
\end{equation}
and satisfy the condition \eref{eq:CScond}.$\blacksquare$

%%%%%%%%%%%%%%%%%%%%%%%%%%%%%%%%%%%%%%%%%%%%%%%%%%%%%%%%%%%%%%%%%%%%%%%%%%%%%%%%%%%%%%%%%%%5

\section{\label{sec:RLDasCE}RLD-based bound as a special case of asymptotic CE bound}

For a channel that admits an RLD-based bound, in order to obtain the
CS condition \eref{eq:RLDcondFin}\textbf{ }in \ref{sec:RLDEquivCS},
we chose $h\!=\!-\frac{\textrm{i}}{2}\nu^{\textrm{AH}}$
that actually satisfies the $\beta_{\tilde{K}}\!=\!0$ constraint \eref{eq:FIC}
of the CE method. This can be verified by taking the $\textrm{Tr}_{\mathcal{H}_{\textrm{out}}}\!\!\left\{ \dots\right\} $
of the both sides of the identity
\begin{equation}
\!\!\!\!\!\!\!\!\!\!\!\!\!\!\!\!\!\!\!\!\!\!\!\!\!\!\!\sum_{ij}h_{ij}\left|K_{j}\right\rangle \!\left\langle K_{i}\right|=\sum_{ij}\frac{\textrm{i}}{2}\left(\nu_{ij}-\nu_{ij}^{\dagger}\right)\left|K_{j}\right\rangle \!\left\langle K_{i}\right|=\frac{\textrm{i}}{2}\sum_{i}\left|\dot{K}_{i}\right\rangle \!\left\langle K_{i}\right|-\left|K_{i}\right\rangle \!\left\langle \dot{K}_{i}\right|\!,
\end{equation}
which results in \eref{eq:FIC}.
% or \eref{eq:RLDcondFin} to obtain directly $\beta_{\tilde{K}}\!=\!0$.
This is consistent, as the CE method must
apply to any $\varphi$-non-extremal channel \cite{Demkowicz2012}
admitting an RLD-based bound. Furthermore, the asymptotic CE bound
\eref{eq:FqCE} is at least as tight as the RLD-based bound \eref{eq:FqChExtRLD}
on the extended channel QFI \eref{eq:FqChExt}. We prove this by substituting
\eref{eq:RLDcondFin} into the definition of $\mathcal{F}^{\textrm{RLD}}\!\left[\Lambda_{\varphi}\otimes\mathcal{I}\right]$
in \eref{eq:FqChExtRLD}, so that
\begin{eqnarray}
\!\!\!\!\!\!\!\!\!\!\!\!\!\!\!\!\!\!\!\!\!\mathcal{F}^{\textrm{RLD}}\!\left[\Lambda_{\varphi}\otimes\mathcal{I}\right] & = & 4\left\Vert \textrm{Tr}_{\mathcal{H}_{\textrm{out}}}\!\!\left\{ \sum_{ij}\nu_{ji}^{\textrm{H}}\left|\tilde{K}_{i}\right\rangle \sum_{pq}\nu_{pq}^{\textrm{H}}\left\langle \tilde{K}_{q}\right|\right\} \right\Vert =4\left\Vert \sum_{i}\dot{\tilde{K}}_{i}^{\dagger}\dot{\tilde{K}}_{i}\right\Vert ,
\end{eqnarray}
where we have used the fact that $\left\langle \tilde{K}_{j}\right|\Omega_{\varphi}^{-1}\left|\tilde{K}_{p}\right\rangle \!=\!\delta_{jp}$.
Hence, $\mathcal{F}^{\textrm{RLD}}\!\left[\Lambda_{\varphi}\otimes\mathcal{I}\right]$
is an example of an asymptotic CE-based bound with a possibly sub-optimal
Kraus representation chosen such that $\forall_{i}:\;\left|\dot{\tilde{K}}_{i}\right\rangle \!=\!\sum_{j}\nu_{ij}^{\textrm{H}}\left|\tilde{K}_{j}\right\rangle $
and $\beta_{\tilde{K}}=0$.$\blacksquare$

%%%%%%%%%%%%%%%%%%%%%%%%%%%%%%%%%%%%%%%%%%%%%%%%%%%%%%%%%%%%%%%%%%%%%%%%%%%%%%%%%%%%%%%%%%%5

\section{\label{sec:QSasCE}Optimal local QS of a channel}

A channel $\Lambda_{\varphi}$ of rank $r$, in order to be locally
\emph{quantum simulable} within small deviations $\delta\varphi$ from a
given $\varphi$, must fulfil the condition (see \Sref{sub:QS})
\begin{eqnarray}
\!\!\!\!\!\!\!\!\!\!\!\!\!\!\!\!\!\!\!\!\!\!\!\!\!\!\!\!\!\!\!\!\!\!\!\Lambda_{\varphi}\!\left[\varrho\right] & = & \textrm{Tr}_{\textrm{E}_{\Phi}\textrm{E}_{\sigma}}\!\!\left\{ U\left(\varrho\otimes\left|\psi_{\varphi}\right\rangle \!\left\langle \psi_{\varphi}\right|\right)U^{\dagger}\right\} +O(\delta\varphi^{2})=\sum_{i=1}^{r^{\prime}\ge r}\bar{K}_{i}(\varphi)\,\varrho\,\bar{K}_{i}(\varphi)^{\dagger}+O(\delta\varphi^{2}),\label{eq:LocQSCond}
\end{eqnarray}
where $\bar{K}_{i}(\varphi)\!=\!\left\langle i\right|U\left|\psi_{\varphi}\right\rangle $
and $\left\{ \left|i\right\rangle \right\} _{i=1}^{r^{\prime}}$ form
any basis in the $r^{\prime}$ dimensional $\mathcal{H}_{\textrm{E}_{\Phi}}\!\times\!\mathcal{H}_{\textrm{E}_{\sigma}}$
space containing $\psi_{\varphi}$. Hence, $\Lambda_{\varphi}$ must
admit a Kraus representation $\{\tilde{K}_{i}\}_{i=1}^{r^{\prime}}$
(with possibly linearly dependent Kraus operators, as for generality
we assume $r^{\prime}\!\ge\! r$) that coincides with the one of \eref{eq:LocQSCond}
up to $O(\delta\varphi^{2})$, i.e. satisfies $\tilde{K}_{i}\!=\!\bar{K}_{i}$
and $\dot{\tilde{K}}_{i}\!=\!\dot{\bar{K}}_{i}$ for all $i$. We
construct a valid decomposition of $|\dot{\psi}_{\varphi}\rangle$ into its
(normalized) components parallel and perpendicular to $\psi_{\varphi}$:
$\left|\dot{\psi}_{\varphi}\right\rangle \!\!=\!\textrm{i}\, a\!\left|\psi_{\varphi}\right\rangle \!-\!\textrm{i}\, b\!\left|\psi_{\varphi}^{\perp}\right\rangle $,
where we can choose $a,b\!\in\!\mathbb{R}$ because of $\partial_{\varphi}\!\left\langle \psi_{\varphi}|\psi_{\varphi}\right\rangle \!=\!0$
and the irrelevance of the global phase. Then, the asymptotic bound
$\mathcal{F}_{\textrm{as}}^{\textrm{bound}}$ of \eref{eq:FqNChAsBound}
determined by the local QS \eref{eq:LocQSCond} at $\varphi$ simply
reads $F_{\textrm{Q}}\!\left[\left|\psi_{\varphi}\right\rangle \right]\!=\!4b^{2}$
and the required Kraus operators $\{\tilde{K}_{i}\}_{i=1}^{r^{\prime}}$
of $\Lambda_{\varphi}$ must fulfil conditions $\tilde{K}_{i}\!=\!\left\langle i\right|U\left|\psi_{\varphi}\right\rangle $
and $\dot{\tilde{K}}_{i}\!=\!\left\langle i\right|U\left|\dot{\psi}_{\varphi}\right\rangle \!\!=\!\textrm{i}\, a\tilde{K}_{i}\!-\!\textrm{i}\, b\left\langle i\right|U\left|\psi_{\varphi}^{\perp}\right\rangle $.
Hence, for the local QS of channel $\Lambda_{\varphi}$ to be valid
$b$ must be finite and we must be always able  to construct
\begin{equation}
U=\left[\begin{array}{ccccc}
\tilde{K}_{1} & \frac{a}{b}\tilde{K}_{1}+\frac{\textrm{i}}{b}\dot{\tilde{K}}_{1} & \bullet & \ldots & \bullet\\
\tilde{K}_{2} & \frac{a}{b}\tilde{K}_{2}+\frac{\textrm{i}}{b}\dot{\tilde{K}}_{2} & \bullet & \ldots & \bullet\\
\tilde{K}_{3} & \frac{a}{b}\tilde{K}_{3}+\frac{\textrm{i}}{b}\dot{\tilde{K}}_{3} & \vdots & \ddots & \vdots\\
\vdots & \vdots & \bullet & \ldots & \bullet
\end{array}\right]\label{eq:UQS}
\end{equation}
with first two columns fixed to give for all $i$ the correct $\left\langle i\right|U\left|\psi_{\varphi}\right\rangle $
and $\left\langle i\right|U\left|\psi_{\varphi}^{\perp}\right\rangle $
respectively. Due to locality, all entries marked with $\bullet$
in \eref{eq:UQS} can be chosen freely to satisfy the unitarity condition
$U^{\dagger}U\!=\! UU^{\dagger}\!=\!\mathbb{I}$. Yet, this constraint
still requires the Kraus operators to simultaneously fulfil $\textrm{i}\sum_{i=1}^{r^{\prime}}\!\dot{\tilde{K}}_{i}^{\dagger}\tilde{K}_{i}\!=\! a\,\mathbb{I}$
and $\sum_{i=1}^{r^{\prime}}\!\dot{\tilde{K}}_{i}^{\dagger}\dot{\tilde{K}}_{i}\!=\!\left(b^{2}+a^{2}\right)\mathbb{I}$.
Without loss of generality, we may shift their phase at $\varphi$,
so that $\tilde{K}_{i}\!\rightarrow\!\textrm{e}^{-\textrm{i}a\varphi}\tilde{K}_{i}$
and the conditions become independent of $a$, i.e. $\textrm{i}\sum_{i=1}^{r^{\prime}}\!\dot{\tilde{K}}_{i}^{\dagger}\tilde{K}_{i}\!=0$
and $\sum_{i=1}^{r^{\prime}}\!\dot{\tilde{K}}_{i}^{\dagger}\dot{\tilde{K}}_{i}\!=\! b^{2}\mathbb{I}$.
Furthermore, these constraints do not require $r^{\prime}\!>\! r$,
as rewriting for example the first one as $\textrm{i}\sum_{i=1}^{r^{\prime}}\!\left\langle \dot{\psi}_{\varphi}\right|U\left|i\right\rangle \left\langle i\right|U\left|\psi_{\varphi}\right\rangle \!=0$,
one can always resolve the identity with some basis vectors $\sum_{i=1}^{r^{\prime}}\!\left|i\right\rangle \,\left\langle i\right|=\sum_{i=1}^{r}\!\left|e_{i}\right\rangle \,\left\langle e_{i}\right|$
and define linearly independent Kraus operators $\left\{ K_{i}\!=\!\left\langle e_{i}\right|U\left|\psi_{\varphi}\right\rangle \right\} _{i=1}^{r}$
also fulfilling the necessary requirements.

Finally, we may conclude that $\Lambda_{\varphi}$ is locally \emph{quantum
simulable} at $\varphi$, if it admits there a Kraus representation
satisfying conditions \eref{eq:QScontrs} stated in the main text,
which due to locality can be generated via \eref{eq:KrausReps} by
some Hermitian $r\!\times\! r$ matrix $h$.$\blacksquare$

%%%%%%%%%%%%%%%%%%%%%%%%%%%%%%%%%%%%%%%%%%%%%%%%%%%%%%%%%%%%%%%%%%%%%%%%%%%%%%%%%%%%%%%%%%%

\section{\label{sec:FinNCEasSDP}Finite-$N$ CE method as a semi-definite programming task}

The finite-N CE bound has been defined in \eref{eq:FqCEN} as

\begin{equation}
\mathcal{F}_{N}^{\textrm{CE}}=4\min_{h}\left\{ \left\Vert \alpha_{\tilde{K}}\right\Vert +\left(N-1\right)\left\Vert \beta_{\tilde{K}}\right\Vert ^{2}\right\} \!,\label{eq:FqCENApp}
\end{equation}
where $\left\Vert \cdot\right\Vert $ denotes the operator norm, $\alpha_{\tilde{K}}\!=\!\sum_{i}\dot{\tilde{K}}_{i}^{\dagger}\dot{\tilde{K}}_{i}$
and $\beta_{\tilde{K}}\!=\!\mathrm{i}\sum_{i}\dot{\tilde{K}}_{i}^{\dagger}\tilde{K}_{i}$.
Given a channel $\Lambda_{\varphi}$ from a $d_{\textrm{in}}$- to
a $d_{\textrm{out}}$-dimensional Hilbert space and the set of its
linearly independent Kraus operators ($d_{\textrm{out}}\!\times\! d_{\textrm{in}}$
matrices) $\left\{ K_{i}\right\} _{i=1}^{r}$, in order to compute
$\mathcal{F}_{N}^{\textrm{CE}}$, we should minimize \eref{eq:FqCENApp}
over locally equivalent Kraus representations \eref{eq:KrausReps}
of $\Lambda_{\varphi}$ generated by all Hermitian, $r\!\times\! r$
matrices $h$.

Basing on the results of \cite{Demkowicz2012}, where the $\beta_{\tilde{K}}\!=\!0$
constraint \eref{eq:FIC} is also imposed on \eref{eq:FqCENApp},
we show that $\mathcal{F}_{N}^{\textrm{CE}}$ can always be evaluated
by means of semi-definite programming (SDP). Adopting a concise notation
in which $\mathbf{K}$ is a column vector containing the starting
Kraus operators $K_{i}$ as its elements, we can associate all locally
equivalent Kraus representations $\tilde{\mathbf{K}}$ in \eref{eq:FqCENApp}
with those generated by any $h$ via $\tilde{\mathbf{K}}\!=\!\mathbf{K}$
and $\dot{\tilde{\mathbf{K}}}\!=\!\dot{\mathbf{K}}\!-\!\mathrm{i}h \mathbf{K}$.
By constructing matrices ($\mathbb{I}_{d}$ represents a $d\!\times\! d$
identity matrix)
\begin{equation}
\!\!\!\!\!\!\!\!\!\!\!\!\mathbf{A}\!=\!\left[\begin{array}{cc}
\sqrt{\lambda_{a}}\mathbb{I}_{d_{\textrm{in}}} & \dot{\tilde{\mathbf{K}}}^{\dagger}\\
\dot{\tilde{\mathbf{K}}} & \sqrt{\lambda_{a}}\mathbb{I}_{r\cdot d_{\textrm{out}}}
\end{array}\right]\quad\quad\quad\mathbf{B}\!=\!\left[\begin{array}{cc}
\sqrt{\lambda_{b}}\mathbb{I}_{d_{\textrm{in}}} & \left(\mathrm{i}\dot{\tilde{\mathbf{K}}}^{\dagger}\mathbf{\tilde{K}}\right)^{\dagger}\\
\mathrm{i}\dot{\tilde{\mathbf{K}}}^{\dagger}\mathbf{\tilde{K}} & \sqrt{\lambda_{b}}\mathbb{I}_{d_{\textrm{in}}}
\end{array}\right]\!,
\end{equation}
which positive semi-definiteness conditions correspond respectively
to
\begin{equation}
\alpha_{\tilde{K}}=\dot{\tilde{\mathbf{K}}}^{\dagger}\dot{\tilde{\mathbf{K}}}\;\leq\,\;\lambda_{a}\mathbb{I}_{d_{\textrm{in}}}\quad\quad\quad\beta_{\tilde{K}}^{\dagger}\beta_{\tilde{K}}=\mathbf{\tilde{K}}^{\dagger}\dot{\tilde{\mathbf{K}}}\,\dot{\tilde{\mathbf{K}}}^{\dagger}\tilde{\mathbf{K}}\;\leq\;\lambda_{b}\mathbb{I}_{d_{\textrm{in}}},
\end{equation}
we rewrite \eref{eq:FqCENApp} into the required SDP form
\begin{eqnarray}
\mathcal{F}_{N}^{\textrm{CE}} & = & 4\min_{h}\left\{ \lambda_{a}+(N-1)\lambda_{b}\right\} ,\label{eq:FqCENAppSDP}\\
 &  & \textrm{s.t. \; }\mathbf{A}\ge0,\,\mathbf{B}\ge0.\nonumber
\end{eqnarray}

For the purpose of this paper we have implemented a semi-definite
program using the CVX package for Matlab \cite{CVX}, which efficiently
evaluates \eref{eq:FqCENAppSDP} given the set of Kraus operators
and their derivatives of a generic channel $\Lambda_{\varphi}$. The
fact that only $\mathbf{K}$ and $\dot{\mathbf{K}}$ are involved
in \eref{eq:FqCENAppSDP} is a consequence of the QFI, and hence the
bound $\mathcal{F}_{N}^{\textrm{CE}}$, being a local quantity.

Lastly, one should note that by slightly modifying the program in
\eref{eq:FqCENAppSDP} we are able to also efficiently evaluate: the
extended channel QFI \eref{eq:FqChExt}, as $\mathcal{F}\!\left[\Lambda_{\varphi}\otimes\mathcal{I}\right]\!=\!\mathcal{F}_{N\!=\!1}^{\textrm{CE}}$;
and the asymptotic extended channel QFI \eref{eq:FqCE}, $\mathcal{F}_{\textrm{as}}\!\left[\Lambda_{\varphi}\otimes\mathcal{I}\right]\!=\!\mathcal{F}_{\textrm{as}}^{\textrm{CE}}$,
by setting $N\!=\!1$ and imposing the $\beta_{\tilde{K}}\!=\!0$
constraint \eref{eq:FIC} as already pursued in \cite{Demkowicz2012}.

%%%%%%%%%%%%%%%%%%%%%%%%%%%%%%%%%%%%%%%%%%%%%%%%%%%%%%%%%%%%%%%%%%%%%%%%%%%%%%%%%%%%%%%%%%%
%%%%%%%%%%%%%%%%%%%%%%%%%%%%%%%%%%%%%%%%%%%%%%%%%%%%%%%%%%%%%%%%%%%%%%%%%%%%%%%%%%%%%%%%%%%
%dummy

\bibliographystyle{iopart-num}
\bibliography{tools_metro}

\end{document}